%% file: DSFprobePaper.tex
\documentclass[aps,pra,twocolumn,superscriptaddress]{revtex4-1}
\input{packages}
\input{definitions}

\begin{document}

\title{Probing the dynamic structure factor of a neutral Fermi superfluid along the BCS-BEC crossover using atomic impurity qubits}

\author{Mark T. Mitchison}
\email{markTmitchison@gmail.com}
\affiliation{Quantum Optics and Laser Science Group, Blackett Laboratory, Imperial College London, London SW7 2BW, United Kingdom}
\affiliation{Clarendon Laboratory, University of Oxford, Parks Road, Oxford OX1 3PU, United Kingdom}

\author{Tomi H. Johnson}
\affiliation{Centre for Quantum Technologies, National University of Singapore, 3 Science Drive 2, 117543 Singapore, Singapore}
\affiliation{Clarendon Laboratory, University of Oxford, Parks Road, Oxford OX1 3PU, United Kingdom}

\author{Dieter Jaksch}
\affiliation{Clarendon Laboratory, University of Oxford, Parks Road, Oxford OX1 3PU, United Kingdom}
\affiliation{Centre for Quantum Technologies, National University of Singapore, 3 Science Drive 2, 117543 Singapore, Singapore}
\affiliation{Keble College, University of Oxford, Parks Road, Oxford OX1 3PG, United Kingdom}

\begin{abstract}
We study an impurity atom trapped by an anharmonic potential, immersed within a cold atomic Fermi gas with attractive interactions that realizes the crossover from a Bardeen-Cooper-Schrieffer (BCS) superfluid to a Bose-Einstein condensate (BEC). Considering the qubit comprising the lowest two vibrational energy eigenstates of the impurity, we demonstrate that its dynamics probes the equilibrium density fluctuations encoded in the dynamic structure factor of the superfluid. Observing the impurity's evolution is thus shown to facilitate nondestructive measurements of the superfluid order parameter and the contact between collective and single-particle excitation spectra. Our setup constitutes a novel model of an open quantum system interacting with a thermal reservoir, the latter supporting both bosonic and fermionic excitations that are also coupled to each other.
\end{abstract}

\maketitle

\section{Introduction}

Dilute gases of cold fermionic atoms with tunable interactions offer a unique platform to study the BCS-BEC crossover~\cite{Greiner2003nat,Zwerger,Randeria2014arcmp}. This fascinating property of Fermi superfluids is broadly relevant for understanding strongly correlated fermion systems, ranging from high-$T_c$ superconductors \cite{Chen2005pr} to quark-gluon plasmas \cite{Schafer2009rpp}. A key observable for probing the physics of the crossover in ultracold gases is their density fluctuations, which reveal the spatial correlations between atoms \cite{Belzig2007pra,Astrakharchik2007pra,Tan2008ap1,*Tan2008ap2,*Tan2008ap3,Kuhnle2010prl,Hoinka2013prl,Hauke2016natphys}, the spectrum of collective excitations \cite{Minguzzi2001epjd,Buchler2004prl,Kinast2004prl,Combescot2006pra,Altmeyer2007prl,Riedl2008pra} and the appearance of bound pairs \cite{Combescot2006epl,Veeravalli2008prl}. However, despite the considerable effort recently devoted to analyzing density fluctuations along the BCS-BEC crossover (for example, Refs.~\cite{Hu2012fp,Guo2013jltp,He2016ap} and references therein), a comprehensive picture is still lacking. 

Density fluctuations in homogeneous systems are characterized by the dynamic structure factor (DSF), denoted by $S(\qq,\nu)$, which plays a central role in the linear-response theory of many-body systems \cite{PinesNozieres}. In a cold-atom setting, $S(\qq,\nu)$ can be measured via Bragg spectroscopy \cite{Stenger1999prl,StamperKurn1999prl,Steinhauer2002prl} or inelastic light scattering \cite{Javanainen1995prl,Csordas1998pra,Cordobes2014prx}. However, these methods have drawbacks, including their destructive nature and diffraction-limited spatial resolution. Accessing the low-$\qq$ regime using Bragg spectroscopy is also challenging, especially in fermionic systems \cite{Hoinka2014PhD}. While measurements of the DSF of a Fermi superfluid are ongoing~\cite{Hoinka2014PhD, Lingham2014prl, Lingham2016jmo}, it is nonetheless worthwhile to consider other techniques for extracting $S(\qq,\nu)$ that do not suffer from the aforementioned shortcomings.

One attractive alternative approach to probing ultracold gases involves monitoring the evolution of an impurity atom embedded within the gas. In this way, it is possible to measure quantities such as temperature \cite{Bruderer2006njp,Sabin2014sr,doubleWellBECThermometer,CrooksTasakiThermometer,Hohmann2016pra}, density \cite{Schmid2010prl,Elliot2016pra} or topological invariants \cite{Grusdt2015}, and to investigate a range of interesting questions bridging quantum optics and many-body physics~\cite{Fedichev2003prl, Astrakharchik2004pra, Klein2005pra, Recati2005prl, Gunter2006prl,   Bruderer2007pra,  Retzker2008prl,  Cirone2009njp, Zipkes2010nature, Haikka2011pra, Will2011prl, Johnson2011pra, Dorner2012if, Catani2012pra, Spethmann2012prl, Scelle2013prl, Balewski2013nature, Wang2015prl, Hohmann2015epjqt, Schmidt2016prl, Johnson2016prl,Jorgensen2016prl,Hu2016prl}. Impurity probes have several advantages over conventional measurement techniques, since a single atom is an easily characterized and controlled system, which may be localized to submicron length scales, and which minimally perturbs its host gas. 

In this article, we propose using an impurity to probe the DSF of a neutral Fermi superfluid manifesting the BCS-BEC crossover. Specifically, we consider an atom trapped by an anharmonic potential, immersed in a weakly confined Fermi gas with tunable attractive interactions. By relating the dissipation rate of the impurity's vibrational energy to $S(\qq,\nu)$, we demonstrate that observing the impurity's evolution allows one to measure the superfluid order parameter and probe the coupling between different kinds of excitations within the gas. Intriguingly, the impurity's environment behaves as a highly tunable phononic thermal reservoir whose excitations may themselves be strongly damped, making our model an unusual open quantum system of intrinsic interest. More generally, we put forward a nondestructive method to probe the dynamic density fluctuations of a degenerate quantum gas, which is valid for arbitrarily strong interactions between the gas atoms, and which can access a spectrum of length scales ranging from the wavelength of collective modes to much less than the interatomic distance.

Impurities in neutral Fermi superfluids have recently been investigated in the context of polaron formation~\cite{Nishida2015prl,Yi2015pra,Lombardi2016arxiv}. However, using impurities to probe ultracold Fermi gases has only been previously discussed in the limit of noninteracting fermions~\cite{Goold2011pra,Knap2012prx,Sindona2013prl}. Our work thus extends the theoretical literature on trapped impurity probes to the setting of fermionic superfluids with tunable interactions. Given the recent surge of experimental progress on impurity immersion in Fermi gases \cite{Spiegelhader2009,Schirotzek2009prl,Ivanov2011prl,Hara2011prl,Hara2014jpsj,Cetina2015prl,Cetina2016sci}, the implementation of our proposal lies within reach.

The remainder of this article is organized as follows. Section~\ref{sec_model} introduces our theoretical model for the Fermi gas, the impurity probe and the dynamics ensuing from their interaction. Section~\ref{sec_spectralFunction} is concerned with general features of the spectral density governing the impurity's evolution, as determined by the density fluctuations of the superfluid. In Section~\ref{sec_results}, we present quantitative results for the impurity decay rate for several examples and use these to illustrate how various properties of the superfluid could be experimentally inferred. We summarize and conclude in Section~\ref{sec_conclusion}.
 
\section{Setup}
\label{sec_model}

The system of interest is an impurity atom of species $A$ and mass $M$, immersed in a cold atomic gas of fermionic species $B$ with mass $m$ and two relevant (hyperfine) spin states. The impurity is confined by a species-selective potential that does not affect the fermions. The trap potential for the fermions is assumed to vary slowly on the length scale of the impurity, so that the gas can be well approximated as homogeneous. In principle, the Fermi gas could be effectively one- or two-dimensional, but we consider only the three-dimensional (3D) case here.

\subsection{Fermi superfluid}

\begin{figure*}
\centering
\begin{minipage}{\linewidth}\centering
\begin{minipage}{0.3\linewidth}
\flushleft(a)\vspace{-5mm}
\begin{figure}[H]\centering
\includegraphics[width=\linewidth]{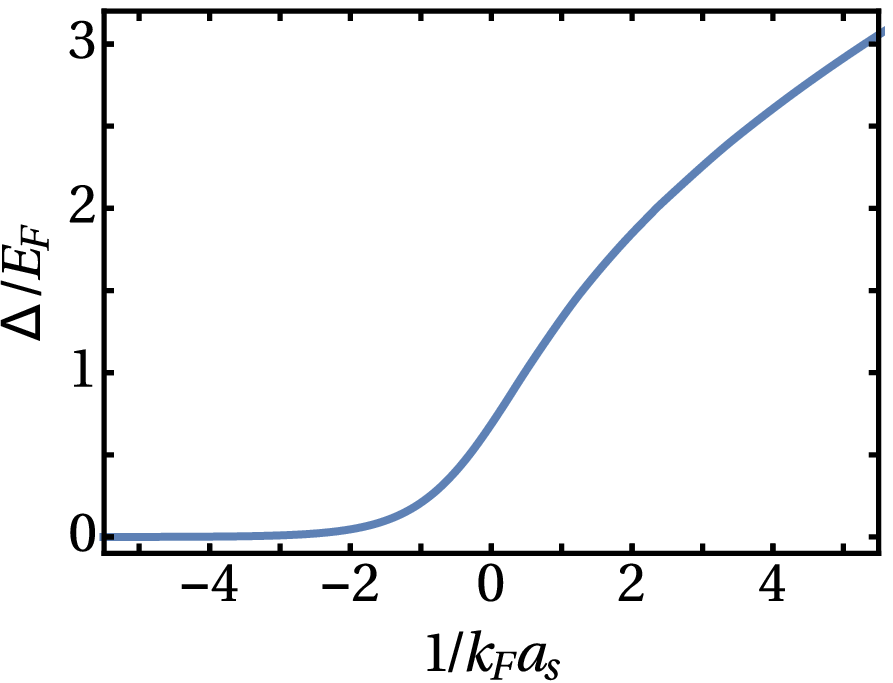}
\end{figure}
\end{minipage}
\begin{minipage}{0.32\linewidth}
\flushleft(b)\vspace{-5mm}
\begin{figure}[H]\centering
\includegraphics[width=\linewidth]{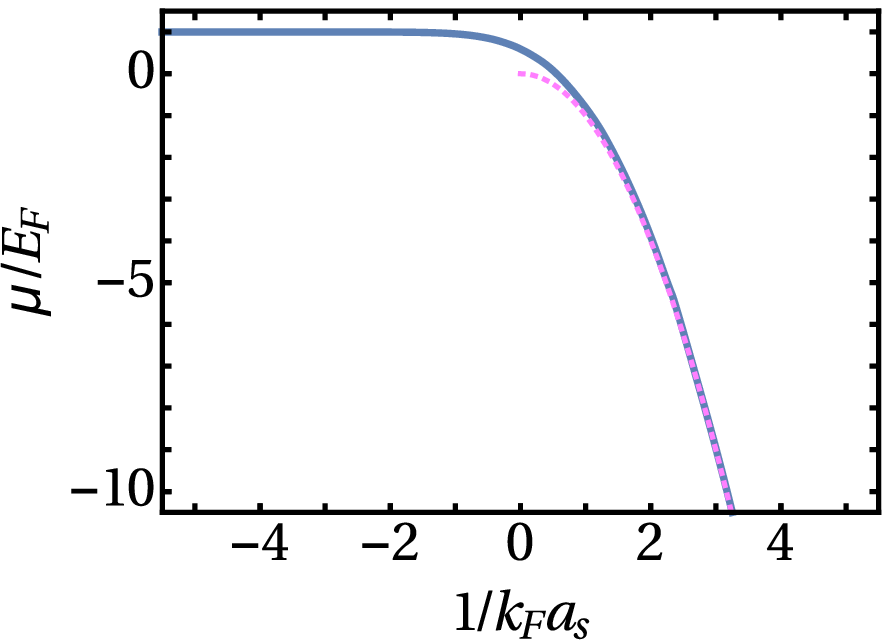}
\end{figure}
\end{minipage}
\begin{minipage}{0.31\linewidth}
\flushleft(c)\vspace{-5mm}
\begin{figure}[H]\centering
\includegraphics[width=\linewidth]{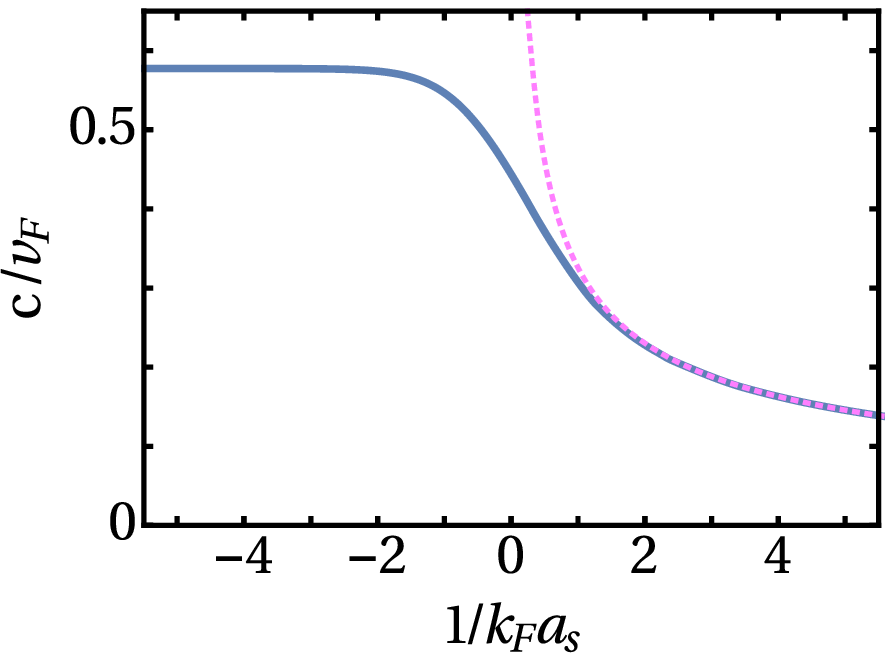}
\end{figure}
\end{minipage}

\end{minipage}
\caption{Evolution of the superfluid's thermodynamic parameters at fixed mean density and zero temperature: (a) order parameter, (b) chemical potential (solid line) shown with the molecular binding energy (dotted line) and (c) speed of sound (solid line) plotted with the corresponding Bogoliubov value for a molecular BEC (dotted line), as described in the main text.\label{DeltaMuC}}
\end{figure*}

The characteristic length and energy scales for a homogeneous 3D Fermi gas with mean number density $\varrho_0$ are fixed by the Fermi wave vector $k_F = (3\pi^2\varrho_0)^{1/3}$ and the Fermi energy $E_F = k_F^2/2m$, where units with $\hbar = 1$ are used throughout. At low temperatures, the dominant interaction between the fermions corresponds to collisions between opposite spins in the $s$-wave channel, described by a scattering length $a_s$ that can be tuned to any value via a Feshbach resonance \cite{Chin2010rmp}. The properties of the system vary markedly as a function of the dimensionless interaction strength $1/k_Fa_s$, ranging from a BCS condensate of Cooper pairs as $1/k_Fa_s \to -\infty$, through the strongly interacting unitary gas at $1/k_Fa_s = 0$, to a BEC of diatomic molecules as $1/k_Fa_s\to +\infty$. 

We use the standard BCS mean-field (saddle-point) approximation to describe the superfluid \cite{Leggett1980jpc,Nozieres1985jltp,deMelo1993prl}. Below the condensation temperature $T_c$, this theory is accurate in the BCS or BEC limits, and is expected to give a qualitatively correct interpolation across the entire crossover~\cite{Engelbrecht1997prb}. The salient aspects of the model are summarized below; further details can be found in  Refs.~\cite{deMelo1993prl,Engelbrecht1997prb,Giorgini2008rmp,BruusFlensberg,Zwerger}, for example. 

The gas is described by the grand-canonical Hamiltonian
\begin{align}
\label{FermiGasHamiltonianRealSpace}
H_B = &\,\sum_{s=\uparrow,\downarrow} \int\dD{3}\rr \;\Psi_s^\d (\rr) \left ( - \frac{\nabla^2 }{2m} - \mu \right ) \Psi_s(\rr) \notag \\ & +\, g\int\dD{3}\rr \; \Psi_\uparrow^\d (\rr) \Psi^\d_\downarrow(\rr)\Psi_\downarrow(\rr) \Psi_\uparrow(\rr).
\end{align}
Here, $\Psi^\d_s(\rr)$ is the atomic field operator which creates a fermion with internal state $s$ at position $\rr$, $\mu$ is the chemical potential, and $g$ is the coupling constant \footnote{The bare coupling constant is expressed in terms of the physical scattering length via the standard prescription~\cite{deMelo1993prl} $ m/4\pi a_s = 1/g + 1/V \sum_\kk^\Lambda m/k^2$, where the summation is cut off at a momentum scale $\Lambda$ satisfying $r_0^{-1} \gg \Lambda \gg a_s^{-1}$, with $r_0$ the range of the interatomic potential. Eliminating $g$ in favour of $a_s$ leads to finite results in the limit $\Lambda\to \infty$; this procedure yields in particular the renormalized gap equation~\eqref{gapEquation}.} of a contact pseudopotential describing interatomic collisions. Below $T_c$, the effective attraction between the fermions leads to the formation of a condensate of Cooper or molecular pairs described by the superfluid order parameter
\begin{align}
\label{deltaDef}
\Delta(\rr) & = g \langle \Psi_\downarrow(\rr) \Psi_\uparrow(\rr) \rangle \notag \\
& = \frac{g}{V} \sum_{\kk,\qq} \ee^{\ii\qq\cdot\rr}  \langle c_{\kk+\qq/2 \downarrow} c_{-\kk+\qq/2 \uparrow} \rangle,
\end{align}
where the angle brackets denote a thermal average. On the second line, the field operators have been expanded in terms of plane-wave mode functions
\begin{equation}
\Psi_s(\rr) = \frac{1}{\sqrt{V}}\sum_\kk \ee^{\ii\kk\cdot\rr} c_{\kk s},
\end{equation}
where $c^\d_{\kk s}$ creates an atom with definite momentum $\kk$ and internal state $s$, and $V$ is a fictitious quantization volume. The translation invariance of the system dictates that $\Delta(\rr) = \Delta$ is spatially constant at equilibrium, meaning that condensation occurs into pair states with zero centre-of-mass momentum, i.e. $\qq=0$. We choose a gauge in which $\Delta$ is real and positive.

The gas supports two distinct kinds of elementary excitation above the ground state. The first kind corresponds to collective oscillations of the pair condensate, which can be resolved into normal modes of wave vector $\qq$ and frequency $\omega_\qq = \omega_q$ (depending only on the magnitude $q = |\qq|$). In the quantum theory, the normal modes are interpreted in terms of phonons carrying momentum $\qq$ and energy $\omega_q$. The phonons have bosonic character since they arise from the collective motion of fermion pairs with centre-of-mass momentum $\qq\neq 0$. This type of excitation has a discrete spectrum, in the sense that each wave vector $\qq$ corresponds to a unique frequency $\omega_q$. 

In the limit $q\to 0$, the phonon dispersion relation becomes sound-like, $\omega_q\to c q$, with $c$ the speed of sound. This manifests the gapless Goldstone mode associated with the $\rm U(1)$ symmetry that is spontaneously broken in the superfluid ground state. Note that a gapless collective mode is a unique feature of the neutral superfluid, since in charged superconductors the Goldstone mode acquires a gap equal to the plasma frequency via the Anderson-Higgs mechanism \cite{Anderson1958pr1,*Anderson1958pr2}. 

Alternatively, if sufficient energy is supplied to the condensate, a bound pair may be broken apart. This creates two fermionic quasiparticles~\footnote{We neglect the possibility of creating a lone quasiparticle by injecting or extracting single atoms into or from the condensate.} with dispersion relation $E_\kk = \sqrt{\Delta^2 + \xi_\kk^2}$, where $\xi_\kk = \varepsilon_\kk -\mu$ and $\varepsilon_\kk = k^2/2m$. Thus, the minimum energy of such a pair excitation is given by the pair gap $\Theta_0 = \min_{\kk,\kk'} (E_{\kk} + E_{\kk'})$, i.e.\
\begin{equation}
\label{pairingGap}
\Theta_0 = \left \lbrace
\begin{array}{ll} 
2\Delta \quad & (\mu \geq 0) \\
2\sqrt{\Delta^2 + \mu^2} \quad & (\mu < 0).
\end{array}\right .
\end{equation}
In general, the energy of a quasiparticle pair carrying centre-of-mass momentum $\qq$ can take any value above the threshold frequency $\Theta_q = \min_\kk ( E_{\kk+\qq/2}+E_{\kk-\qq/2})$. Thus, the spectrum of pair excitations carrying momentum $\qq$ is continuous.

From here on, our analysis is restricted to temperatures well below the pair excitation gap, i.e.\ $\beta\Theta_0\gg 1$, where $\beta = 1/k_B T$ is the inverse temperature of the gas, so that the number of thermally excited fermionic quasiparticles is negligibly small. In this case, the order parameter and chemical potential are determined by the equations of state~\cite{Engelbrecht1997prb}
\begin{align}
\label{gapEquation}
\frac{m}{4\pi   a_s} & = \frac{1}{V}\sum_{\kk}\left ( \frac{1}{2\varepsilon_\kk}  - \frac{1}{2E_{\kk}}\right ), \\
\label{densityEquation}
\varrho_0 & =  \frac{1}{V}\sum_{\kk}\left ( 1 - \frac{\xi_\kk}{E_\kk} \right ),
\end{align}
while the speed of sound is given by \cite{Combescot2006pra,Zhang2011pra}
\begin{equation}
\label{soundSpeed}
c^2 = \frac{1}{3m^2}\frac{\Delta^2 J_2J_4}{\Delta^2J_2^2 + J_\xi^2},
\end{equation}
where $J_\xi = J_4 - \mu J_2$ and
\begin{equation}
\label{J2}
J_2 = \frac{1}{V}\sum_\kk \frac{1}{E_\kk^3},\quad J_4 = \frac{1}{V}\sum_\kk \frac{k^2}{E_\kk^3}.
\end{equation} 

For reference, the well-known evolution of the order parameter, chemical potential and sound speed as a function of $1/k_Fa_s$ is plotted in Fig.~\ref{DeltaMuC} for several values in the vicinity of the crossover. In the BCS limit, the system can be understood as a condensate of weakly bound Cooper pairs residing close to the Fermi surface, such that $\Delta\ll E_F$, $\mu = E_F$ and $c = v_F/\sqrt{3}$ is the Bogoliubov-Anderson sound speed \cite{Anderson1958pr1,*Anderson1958pr2}, where $v_F = k_F/m$ is the Fermi velocity. On the other hand, in the BEC limit almost all atoms are paired into tightly bound diatomic molecules, leading to $\Delta \gg E_F$, with $c = \sqrt{\pi a_s \varrho_0}/m = \sqrt{4\pi a_b \varrho_b}/m_b$ the Bogoliubov sound speed in a condensate of bosons having mass $m_b=2m$, density $\varrho_b=\varrho_0/2$ and scattering length $a_b = 2a_s$ (this mean-field result differs from the more accurate value $a_b \approx 0.6 a_s$ \cite{Petrov2004prl}), while $\mu= -1/(2ma_s^2)$ is the molecular binding energy.

\subsection{Impurity probe}

\begin{figure}
\centering
\begin{minipage}{\linewidth}\centering

\begin{minipage}{0.49\linewidth}
\flushleft(a)\vspace{-3mm}
\begin{figure}[H]\centering
\includegraphics[width=0.7\linewidth]{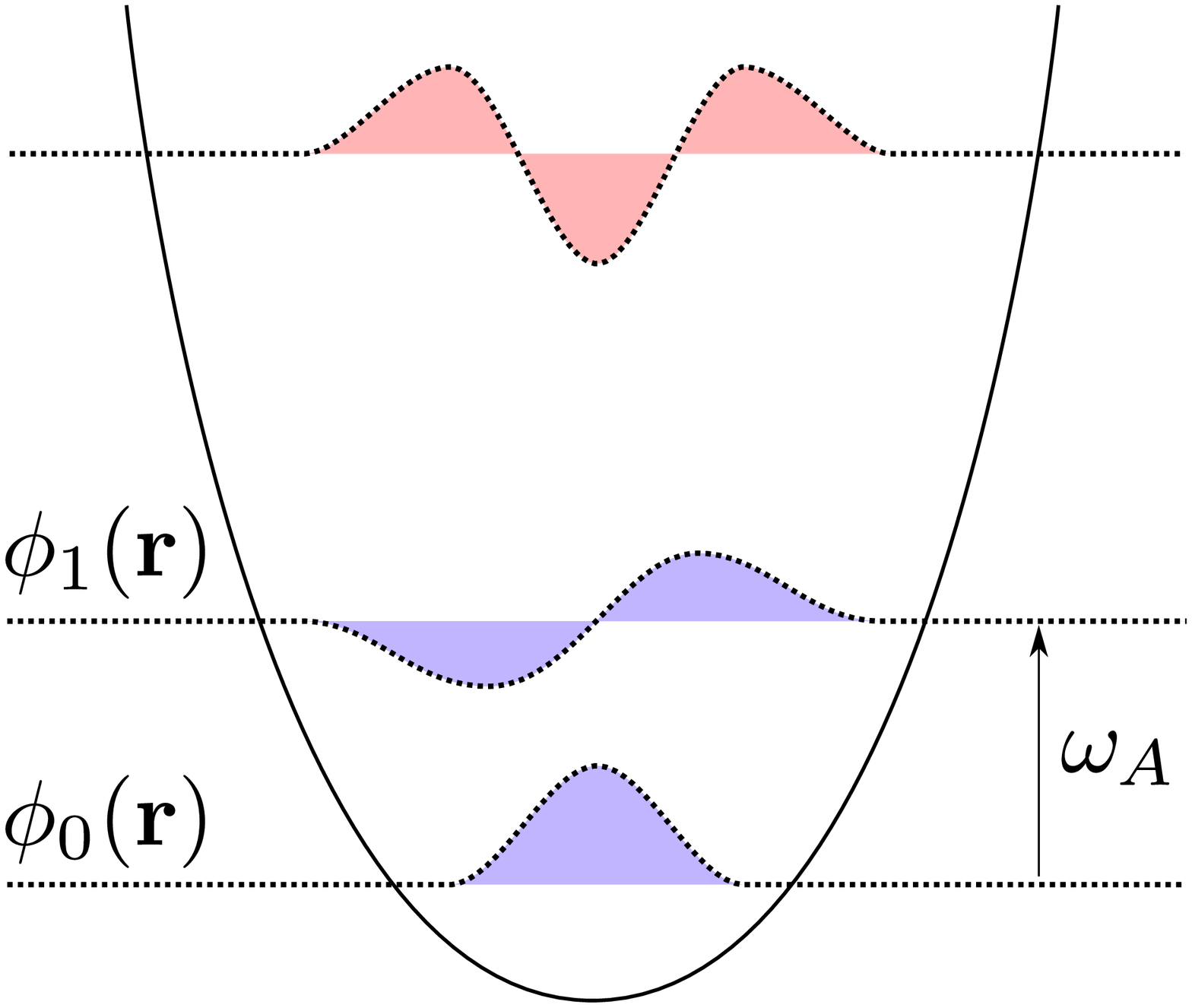}
\end{figure}
\end{minipage}
\begin{minipage}{0.49\linewidth}
\flushleft(b)\vspace{-4mm}
\begin{figure}[H]\centering
\includegraphics[width=\linewidth]{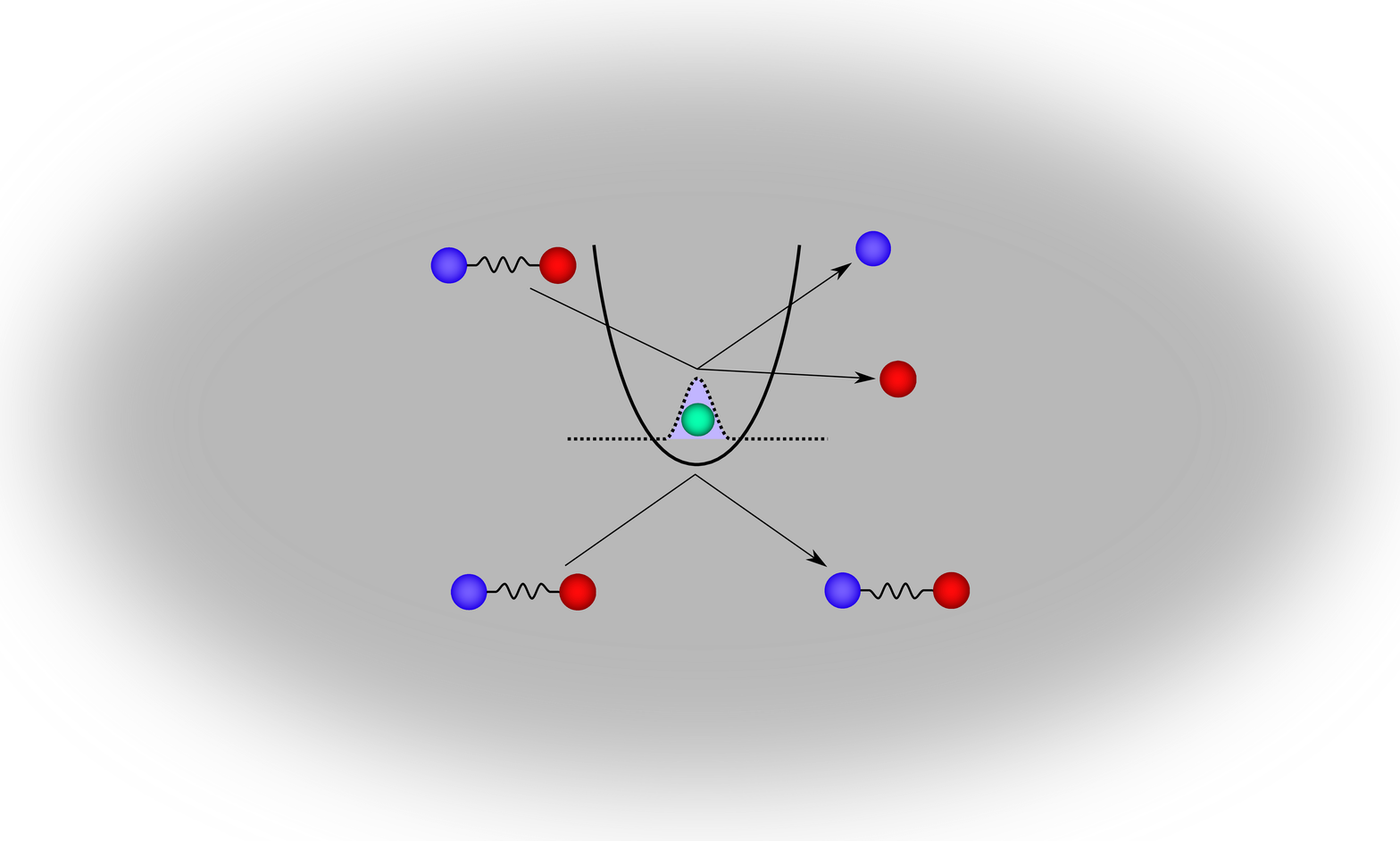}
\end{figure}
\end{minipage}
\end{minipage}
\caption{Schematic depiction of the probe. (a)~Level scheme of an impurity qubit comprising the lowest two vibrational eigenstates $\phi_{0}(\rr)$ and $\phi_1(\rr)$ of an anharmonic potential, which are well separated in energy from higher excited states. (b)~Scattering of gas atoms by the impurity may either impart momentum to condensed pairs, thus exciting the collective mode, or break pairs apart into two fermionic quasiparticles.\label{impurityCartoon}}
\end{figure}

The probe consists of an impurity atom confined by a strongly anharmonic potential, such that the lowest two energy levels of its centre-of-mass motion are well separated in energy from the rest. We can thus restrict our attention to the qubit comprising these states, $\ket{0}$ and $\ket{1}$, described by the Hamiltonian
\begin{equation}
\label{HimpurityTwoLevel}
H_A = \omega_A \sigma^\d\sigma,
\end{equation}
where $\sigma = \ketbra{0}{1}$ and $\omega_A/2\pi$ is the frequency of small oscillations close to the potential minimum. All other motional and internal states of the impurity are assumed to be far from resonance or otherwise negligible. The level structure is depicted schematically in Fig.~\ref{impurityCartoon}(a).

We suppose that impurity-gas collisions are elastic and independent of spin. This assumption may hold approximately \cite{Spiegelhader2009,Cetina2015prl}, or as an exact consequence of rotation symmetry if the impurity is spinless and the internal states of the atoms comprising the host gas have the same total hyperfine spin \cite{Ivanov2011prl,Hara2011prl}. The coupling between the impurity and the gas is modelled using an $s$-wave contact pseudopotential, which in the qubit subspace reads as
\begin{align}
\label{HabRealSpace}
H_{AB}  = \kappa\sum_{i,j= 0,1}\ketbra{i}{j} \int\dD{3}\rr\; \phi^*_i(\rr)\phi_j(\rr) \varrho(\rr),
\end{align}
where $\phi_{i}(\rr) = \braket{\rr}{i}$, for $i = 0,1$, are the impurity energy eigenfunctions, $\varrho(\rr) = \sum_s\Psi_s^\d(\rr) \Psi_s(\rr)$ is the fermion number density, while the coupling constant is $\kappa = 2\pi \bar{a}/\bar{m}$, with $\bar{a}$ the interspecies $s$-wave scattering length and $\bar{m} = m M /( m + M)$ the reduced mass. 

The off-diagonal terms in Eq.~\eqref{HabRealSpace} with $i \neq j$ describe the dissipation of vibrational energy, while the diagonal terms with $i=j$ lead to dephasing. As shown in Appendix \ref{app_superfluidME}, only the off-diagonal terms contribute to the dynamics of the impurity (at lowest nontrivial order). The diagonal contributions can be neglected because the rate of dephasing vanishes in a 3D superfluid environment, even at finite temperature \cite{Bruderer2006njp}. Introducing the Fourier transform of the density
\begin{align}
\label{densityFourier}
\varrho_{\qq} & = \int\dD{3}\rr\; \ee^{-\ii\qq\cdot\rr} \varrho(\rr) \notag \\ 
& = \sum_{s=\uparrow,\downarrow}\sum_\kk c_{\kk-\qq/2,s}^\d c_{\kk+\qq/2,s},
\end{align}
and the coupling constants 
\begin{equation}
\label{lambdaMuNu}
\lambda_\qq = \kappa \int\dD{3}\rr \; \ee^{\ii\qq\cdot\rr} \phi^*_1(\rr)\phi_0(\rr),
\end{equation}
we  rewrite the impurity-gas interaction in the suggestive form
\begin{align}
\label{HabApproxQspace}
H_{AB}  = \frac{1}{V}\sum_{\qq} \left ( \lambda_{\qq}\sigma^\d \varrho_\qq + \lambda_\qq^* \sigma \varrho_\qq^\d \right ).
\end{align}

The operators $\varrho_\qq$ ($\varrho^\d_\qq$) annihilate (create) excitations of momentum $\qq$ in the gas, without changing the number of atoms~\footnote{Formally, this property is summarized by the commutation relations $[{\bf P},\varrho_\qq] = -\qq \varrho_\qq$ and $[N_B,\varrho_\qq] = 0$, where $\bf P$ and $N_B$ respectively denote the centre-of-mass momentum and the total atom number of the Fermi gas}. This leads to a physical interpretation of Eq.~\eqref{HabApproxQspace} in terms of the production and destruction of density excitations by the impurity, over a range of momenta fixed by the coupling constants $\lambda_\qq$. The situation is reminiscent of atom-photon interactions in quantum electrodynamics, where changes of the atomic state are associated with the emission and absorption of electromagnetic quanta. In the present case, the impurity may create and destroy either phonons or fermionic pair excitations, as illustrated in Fig.~\ref{impurityCartoon}(b).

\subsection{Master equation}
\label{sec_fermiME}

We now show that the impurity dynamics probes the spectrum of density fluctuations in the gas. To that end, we use a Lindblad master equation to describe the evolution of the impurity's reduced density matrix $\rho_A(t) = \Tr_B [\rho(t)]$, where $\rho(t)$ denotes the global quantum state at time $t$. The master equation is derived in Appendix~\ref{app_superfluidME} under the Born-Markov and rotating-wave approximations \cite{BreuerPetruccione}, assuming that the coupling between the impurity and the gas is weak and that initial correlations between them can be neglected. Specifically, we take an initial product state $\rho(0) = \rho_A(0) \rho_B$, where $\rho_B =\ee^{-\beta H_B}/\ZZ_B$, with $\ZZ_B = \Tr[\ee^{-\beta H_B}]$ the partition function. The environment-induced renormalization of the impurity vibrational frequency is absorbed into the definition of $\omega_A$. We also assume that $\beta \omega_A \gg 1$.

The master equation obtained under the foregoing assumptions may be written as
\begin{equation}
\label{fermiSuperfluidME}
\dt{\rho_A} = \ii [\rho_A,H_A] +  \Gamma \left ( \sigma \rho_A \sigma^\d - \frac{1}{2} \left \lbrace \sigma^\d \sigma, \rho_A \right \rbrace \right ).
\end{equation}
This equation describes the loss of vibrational energy by the impurity at a rate $\Gamma = 2\pi \SI(\omega_A)$, where we defined the spectral density of the fermionic environment as
\begin{equation}
\label{generalisedSpectralDensity}
\SI(\nu) = \frac{1}{V}\sum_{\qq}\lvert \lambda_\qq\rvert^2 S(\qq,\nu),
\end{equation}
which is expressed in terms of the DSF
\begin{equation}
\label{DSFdefinition}
S(\qq,\nu) = \frac{1}{2\pi}\int_{-\infty}^\infty\dd t \int\dD{3}\rr\; \ee^{\ii( \nu t - \qq\cdot\rr)}\mathcal{C}(\rr,t).
\end{equation}
Here, we introduced the autocorrelation function of the density fluctuations
\begin{equation}
\label{densityFluctuationAutocorrelator}
\mathcal{C}(\rr-\rr',t-t') = \left \langle \delta \varrho(\rr,t) \delta \varrho(\rr',t')\right \rangle,
\end{equation}
with $\delta\varrho(\rr,t) = \varrho(\rr,t) - \varrho_0$ and $\varrho(\rr,t) = \ee^{\ii H_B t}\varrho(\rr)\ee^{-\ii H_B t}$.  Equations~\eqref{fermiSuperfluidME}--\eqref{densityFluctuationAutocorrelator} establish a direct connection between the dissipative dynamics of the impurity and the density fluctuations of the superfluid. Note that the DSF is temperature-dependent --- in particular, it satisfies detailed balance $S(\qq,-\nu) = \ee^{-\beta\nu} S(\qq,\nu)$ --- so our definition of the spectral density also depends on temperature. 

The decay rate $\Gamma$ can be measured by observing the evolution of the population of the excited state, which undergoes pure exponential decay in time according to  $p_1(t) = \matrixel{1}{\rho_A(t)}{1} = \ee^{-\Gamma t}p_1(0)$. Varying the trap frequency from one measurement to the next enables reconstruction of $\Gamma$ as a function of $\omega_A$ over many experimental runs, from which properties of the DSF can be inferred due to the relation~\eqref{generalisedSpectralDensity}, as we explain in detail in subsequent sections. 

The measurement signal from each experiment can be increased by simultaneously observing many impurities immersed within a single realization of the Fermi gas, so long as these impurities remain uncorrelated. In Appendix~\ref{app_superfluidME}, we show how to configure the impurities so that they evolve independently, despite their mutual interaction with the gas. This is possible because the impurities behave like acoustic dipoles, which emit and absorb density waves anisotropically \cite{Russell1999amjp}. Therefore, the impurities can be arranged so that the phonon radiation emitted by each one is not absorbed by the others. Similar arguments were used in our previous work on thermometry using impurities \cite{doubleWellBECThermometer}. 

We conclude this subsection by briefly reviewing the approximations underlying the master equation \eqref{fermiSuperfluidME}. Our basic assumption is that the environment correlation time $\tau_B$ is the shortest time scale of the problem. We can approximate the correlation time by $\tau_B \approx \ell/c$, which estimates the time taken for a spontaneously emitted phonon to irreversibly propagate away from the impurity's domain of influence (see Section~\ref{sec_phononRegimeFermi}). The Born-Markov approximation then requires that $\Gamma \ll c/\ell$, while the rotating-wave approximation is valid when $\Gamma\ll \omega_A$. Both of these inequalities are well satisfied in all the examples that follow. 

\subsection{Form factor in the harmonic, isotropic approximation}

Isotropy of the superfluid implies that $S(\qq,\nu) = S(q,\nu)$ is independent of the direction of $\qq$. Working in the thermodynamic limit via the substitution $(2\pi)^3 \sum_\qq \to V\int\dD{3}\qq $, we find from Eq.~\eqref{generalisedSpectralDensity} that
\begin{equation}
\label{spectralDensityAngularIntegrated}
\SI(\nu) = \frac{\kappa^2}{2\pi^2}\int_0^\infty\dd q\; q^2 \Phi(q) S(q,\nu),
\end{equation}
where we defined the dimensionless form factor $\Phi(q)$ by an angular average of the squared coupling constants, i.e.
\begin{equation}
\label{formFactor}
\Phi(q) = \frac{1}{4\pi\kappa^2} \int_{S^2}\dd\Sigma_\qq \; \vert \lambda_\qq\vert^2,
\end{equation}
with $\dd\Sigma_\qq$ the surface element on the unit sphere $S^2$ in $\qq$-space. The form factor encapsulates the effect of the impurity's geometry on its dissipative dynamics.

In order to explicitly evaluate the form factor, we choose a simple approximation for the impurity potential and wave functions enabling us to gain analytical insight into the problem. Specifically, we assume that the impurity potential is isotropic, and use the eigenfunctions of the simple harmonic oscillator, with the excited state corresponding to motion in the $z$ direction (these approximations are justified in the following paragraph). Neglecting a normalization factor, the chosen wave functions are $\phi_i(\rr) \propto H_i(z/\ell) \ee^{-r^2/2\ell^2}$, for $i= 0,1$, where $H_i(z/\ell)$ is the $i^{\mathrm{th}}$ Hermite polynomial as a function of the $z$ coordinate expressed in units of the natural oscillator length $\ell = \sqrt{1/M\omega_A}$. The form factor in this approximation is given by
\begin{equation}
\label{formFactorSHO}
\Phi(q) = \frac{1}{6}\ell^2 q^2 \ee^{-\ell^2 q^2/2}.
\end{equation}

Let us now discuss the validity of the approximations leading to Eq.~\eqref{formFactorSHO} and their compatibility with the qubit representation of the impurity. The harmonic approximation correctly captures the symmetry of the impurity eigenfunctions for a parity-invariant potential, yet neglects distortions due to the necessary anharmonicity of the potential. Quantitative corrections may therefore be needed for strongly anharmonic potentials. In addition, the first excited state of an isotropic potential is triply degenerate, apparently conflicting with the two-level approximation for the impurity. However, the additional degenerate sublevels can be consistently neglected, because the rate of environment-induced transitions between the excited sublevels vanishes under the Born-Markov approximation, due to the super-Ohmic nature of the reservoir (see Section~\ref{sec_phononRegimeFermi} and Appendix~\ref{app_superfluidME}). This conclusion remains true even at finite temperature, so long as $\beta\Theta_0\gg 1$. 

We emphasize that the assumption of an isotropic potential and harmonic eigenfunctions is inessential. Indeed, detailed knowledge of the impurity wave functions is not necessary for the implementation of our proposal.

\section{Dynamic structure factor and spectral density}
\label{sec_spectralFunction}

In this section, we describe the general properties of the spectral density, which follow from the frequency and momentum dependence of the DSF. We obtain the latter from the linear susceptibility $\chi(q,\nu)$, whose imaginary part is related to $S(q,\nu)$ by the fluctuation-dissipation theorem \cite{PinesNozieres}
\begin{equation}
\label{fluctuationDissipationTheorem}
S(q,\nu) = \frac{-1}{\pi(1-\ee^{-\beta\nu})}\Im\left [\chi(q,\nu+\ii\epsilon)\right ],
\end{equation}
where $\epsilon$ is a positive infinitesimal. 

Identical expressions for the susceptibility in BCS theory have been obtained by various authors using several formally different methods, including kinetic equations~\cite{Combescot2006pra,Guo2013ijmpb,Guo2013jltp}, the random-phase approximation \cite{Cote1993prb,Zou2010pra} and functional integrals~\cite{He2016ap}. Importantly, all these calculations self-consistently account for fluctuations of the order parameter resulting from local perturbations of the density. Incorporating the dynamics of the order parameter is necessary in order to preserve gauge invariance and to recover the contribution of the collective mode \cite{Anderson1958pr1,*Anderson1958pr2}. In addition, this result for $\chi(q,\nu)$ agrees well with experimental Bragg spectroscopy data for $q\gg k_F$ \cite{Zou2010pra}. 

In the following subsections, we separately discuss the qualitative features of the spectral density and the DSF at frequencies below and above the pair gap. Explicit expressions for $S(q,\nu)$ and $\chi(q,\nu)$ can be found in Appendix \ref{app_DSFderivation}. 

\subsection{Frequencies below the pair gap}
\label{sec_phononRegimeFermi}

First we consider frequencies smaller than the pair gap, $|\nu| < \Theta_0$. Here, the DSF is given by
\begin{equation}
\label{DSFcolective}
S(q,\nu) = W_q \left [ (1 + n_q)\delta(\nu - \omega_q) + n_q \delta(\nu + \omega_q) \right ],
\end{equation}
where $n_q = n(\omega_q) = (\ee^{\beta\omega_q} - 1)^{-1}$ is the Bose-Einstein distribution. This describes the possibility to absorb or emit a collective mode excitation (phonon) with wave vector $\qq$ and frequency $\omega_q$. The spectral weight $W_q$ and dispersion relation $\omega_q$ are given by complicated expressions that must be computed numerically, in general. 

The spectral density for $|\nu|< \Theta_0$ is thus of the form
\begin{equation}
\label{spectralFunctionFermiPhonon}
\SI(\nu) = \left \lbrace \begin{array}{ll} 
\SJ(\nu)[1 + n(\nu)] & \quad (\nu > 0)\\
\SJ(|\nu|)n(|\nu|) & \quad (\nu < 0),
\end{array}\right. 
\end{equation}
where the spectral density at zero temperature is 
\begin{equation}
\label{spectralDensityFermi}
\SJ(\nu) = \sum_\qq \lvert \tilde{\lambda}_\qq\rvert^2 \delta(\nu - \omega_q),
\end{equation}
with the rescaled coupling $\tilde{\lambda}_\qq = \lambda_\qq \sqrt{W_q/V}$. Note that Eqs.~\eqref{spectralFunctionFermiPhonon} and \eqref{spectralDensityFermi} take the form of the spectral density for a linear, harmonic, bosonic environment (see, for example, Ref.~\cite{doubleWellBECThermometer}). Therefore, at low frequencies the fermionic superfluid behaves identically to a bosonic reservoir (at this level of approximation).

As $q\to 0$, the collective mode dispersion relation is linear, $\omega_q \approx c q$, while the spectral weight is approximately $W_q \approx \varrho_0\varepsilon_q/\omega_q$ (recall that $\varepsilon_q = q^2/2m$). These approximations hold for wave vectors $\zeta q \ll 1$, where $\zeta = c/\Delta$ is the coherence length \cite{Buchler2004prl}. In this regime, the DSF exhausts the $f$-sum and compressibility sum rules \cite{PinesNozieres}, meaning that the collective mode is the only relevant long-wavelength excitation at \textit{any} frequency. 

Now, if we consider impurity potentials such that $\ell \gg \zeta$, the form factor \eqref{formFactorSHO} samples only wave vectors in the range $\zeta q\ll 1$. In such a case, Eq.~\eqref{spectralFunctionFermiPhonon} is valid at all frequencies, with the zero-temperature spectral density given by the super-Ohmic form
\begin{equation}
\label{superOhmicSpectralDensityFermi}
\SJ(\nu)  = \alpha \omega_c^{-4} \nu^5 \ee^{-\nu^2/2\omega_c^2},
\end{equation}
where we defined the dimensionless coupling strength $\alpha = \kappa^2 \varrho_0/(24\pi^2 m \ell^2 c^3)$ and the frequency cutoff $\omega_c = c/\ell$. In this regime, the environment correlation time (the inverse of its frequency bandwidth) is $\tau_B = \omega_c^{-1}$, in agreement with the estimate provided in Section \ref{sec_fermiME}.  

In the general case, Eq.~\eqref{superOhmicSpectralDensityFermi} is valid only in the limit $\nu\to 0$. At finite frequencies satisfying $|\nu|<\Theta_0$, the spectral density is determined by
\begin{equation}
\label{phononSpectralDensity}
\SJ(\nu) = \kappa^2\Phi(q_\nu) W_{q_\nu} D(\nu),
\end{equation}
where the wave vector $q_\nu$ is defined by $\omega_{q_\nu} = \nu$ and we introduced the phonon density of states
\begin{equation}
\label{phononDensityOfStates}
D(\nu) = \frac{1}{V}\sum_{\qq} \delta(\nu - \omega_q).
\end{equation}

The spectral density \eqref{phononSpectralDensity} arises from three factors. The first term $\kappa^2\Phi(q)$ is geometrical in origin, describing how the impurity's density profile affects the transfer of momentum to scattered gas atoms. In order to interpret the weight $W_q$, we insert a complete set of energy eigenstates into Eq.~\eqref{DSFdefinition} to obtain, at zero temperature,
\begin{equation}
\label{DSFkallenLehmann}
S(q,\nu) = \frac{1}{V}\sum_n \left \lvert \matrixel{\Omega_n}{\varrho_\qq^\d}{\Omega_0}\right \rvert^2\delta(\nu -\Omega_n),
\end{equation}
where the sum runs over all eigenstates $\ket{\Omega_n}$ of $H_B$ having energy $\Omega_n$ above the ground state $\ket{\Omega_0}$. Comparison of Eqs.~\eqref{DSFcolective} and \eqref{DSFkallenLehmann} leads to the heuristic identification $W_q \sim \lvert \bra{1_\qq}\varrho_\qq^\d \ket{\Omega_0} \rvert^2/V$, where $\ket{1_\qq}$ represents an excited state occupied by a single phonon with momentum $\qq$. Thus, $W_q$ quantifies the ease with which a collective excitation can be created from the ground state by single-particle scattering events described by the operator $\varrho_\qq^\d$. The final factor entering Eq.~\eqref{phononSpectralDensity} gives the density of single-phonon states $D(\nu)$ available to be excited by such a process. 

\subsection{Frequencies above the pair gap}
\begin{figure}
\centering
\begin{minipage}{\linewidth}\centering

\begin{minipage}{0.49\linewidth}
\flushleft(a)\vspace{-7mm}
\begin{figure}[H]\centering
\includegraphics[width=0.7\linewidth]{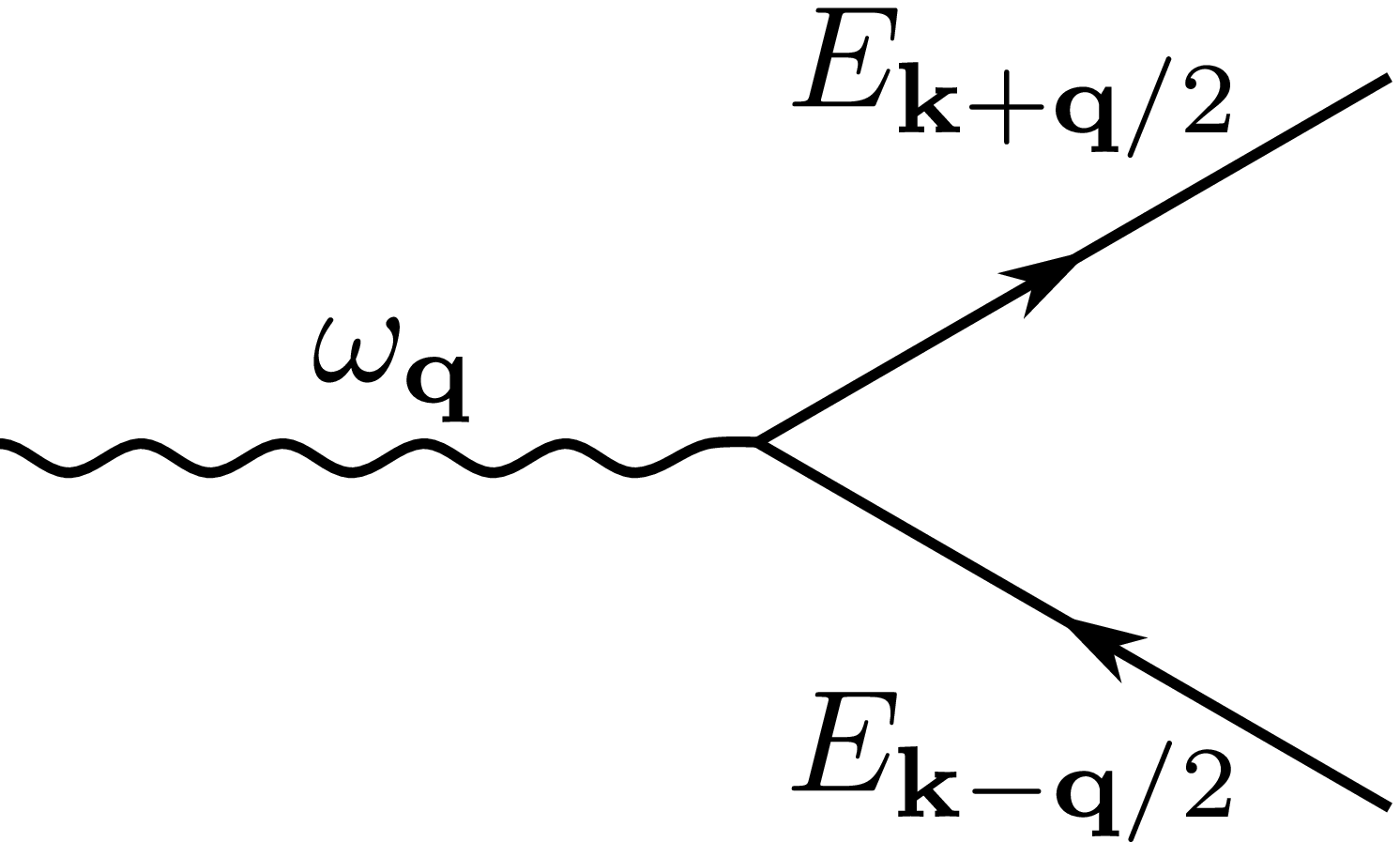}
\end{figure}
\end{minipage}
\begin{minipage}{0.49\linewidth}
\flushleft(b)\vspace{-7mm}
\begin{figure}[H]\centering
\includegraphics[width=0.7\linewidth]{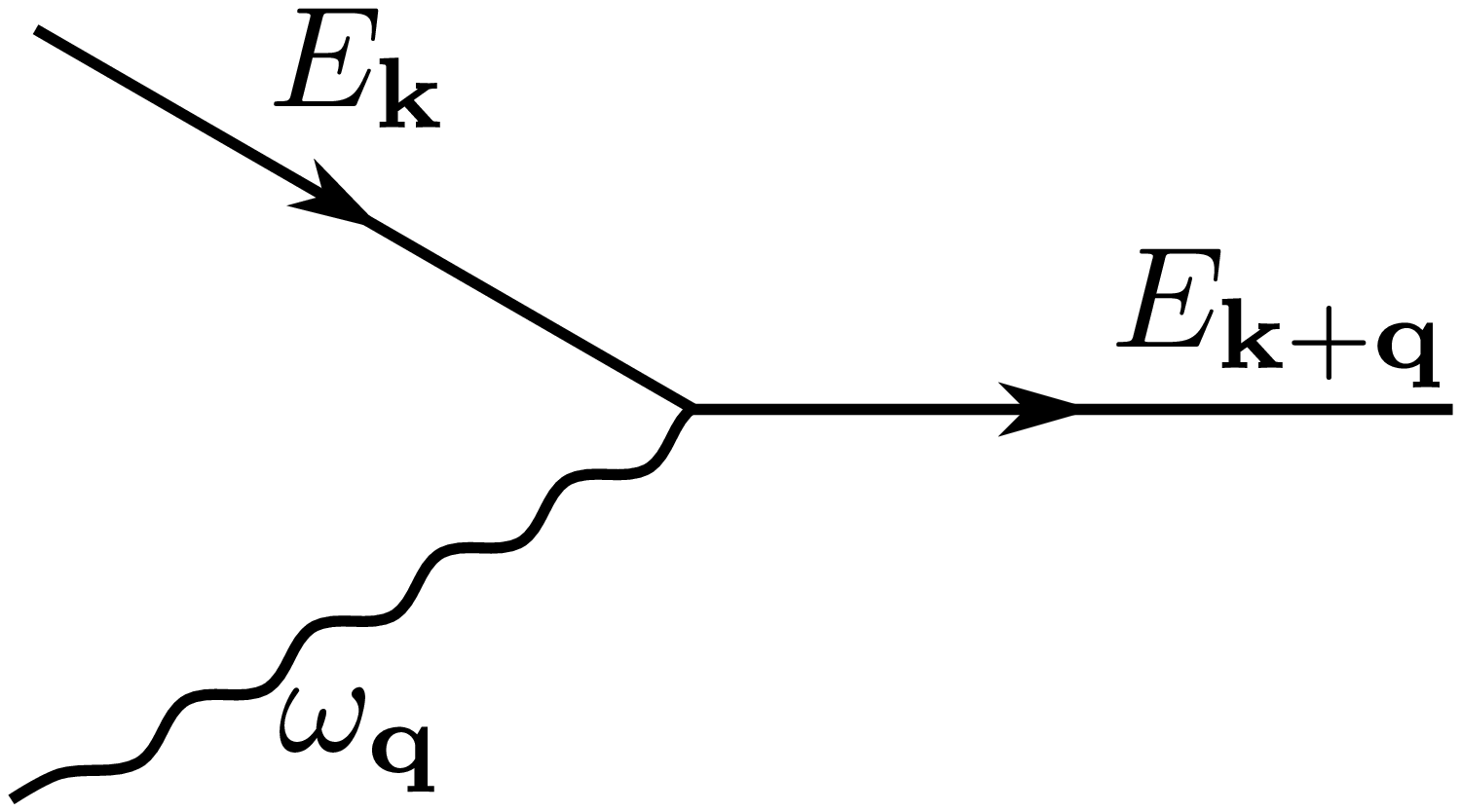}
\end{figure}
\end{minipage}
\end{minipage}
\caption{Diagrammatic representation of (a) Beliaev and (b) Landau decay due to the coupling between phonons (wiggly lines) and fermionic quasiparticles (straight lines), with time increasing from left to right.\label{BeliaevLandau}}
\end{figure}

At frequencies larger than the pair gap, the DSF cannot be written in a simple form, thus we give here only a qualitative description. For $\nu>\Theta_0$, there is sufficient energy to break apart condensed pairs into two fermionic quasiparticles. Such processes contribute to $S(q,\nu)$ over all frequencies $\nu > \Theta_q$, corresponding to the continuum of fermionic pair excitation modes at momentum $\qq$. The collective mode response is also broadened into a continuum at high frequencies, because phonons with energy $\omega_q > \Theta_q$ can decay into a pair of fermionic quasiparticles, as shown diagrammatically in Fig.~\ref{BeliaevLandau}(a). This so-called Beliaev damping leads to a finite lifetime for each phonon mode carrying momentum $\qq$, with a corresponding frequency uncertainty around the resonance at $\omega_q$. The importance of Beliaev damping depends on the value of $1/k_Fa_s$, as discussed in Section~\ref{sec_separationOfSpectra}.

\begin{figure*}[t]\centering

  \begin{minipage}{\linewidth}\centering
     \begin{minipage}{0.3\linewidth}\centering
          \flushleft (a) \vspace{-8mm} \begin{figure}[H]\centering
			\includegraphics[width=0.9\linewidth]{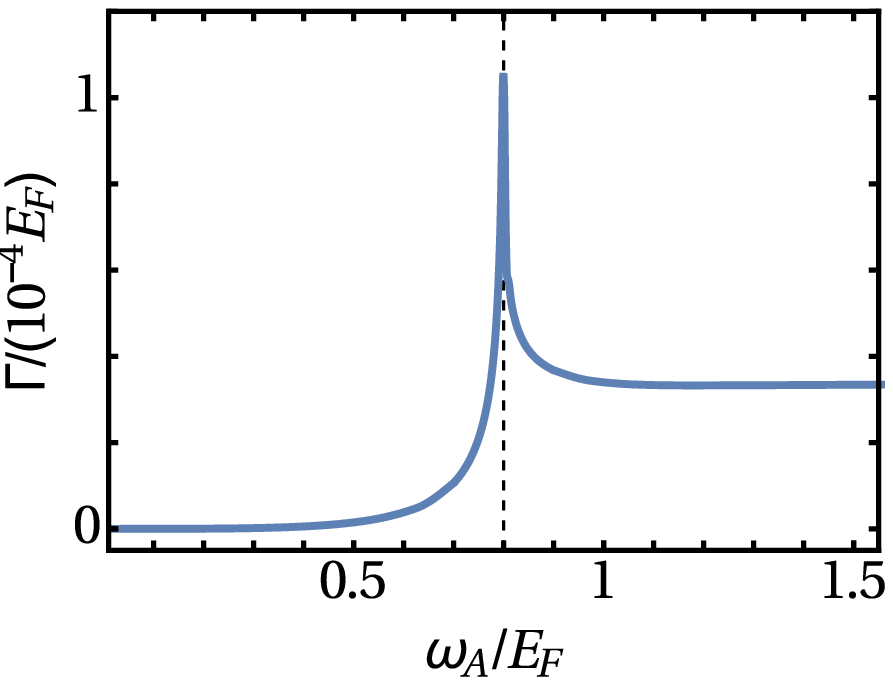}
          \end{figure}
      \end{minipage}
      \begin{minipage}{0.3\linewidth}\centering
          \flushleft (b) \vspace{-8mm} \begin{figure}[H]\centering
              \includegraphics[width=0.9\linewidth]{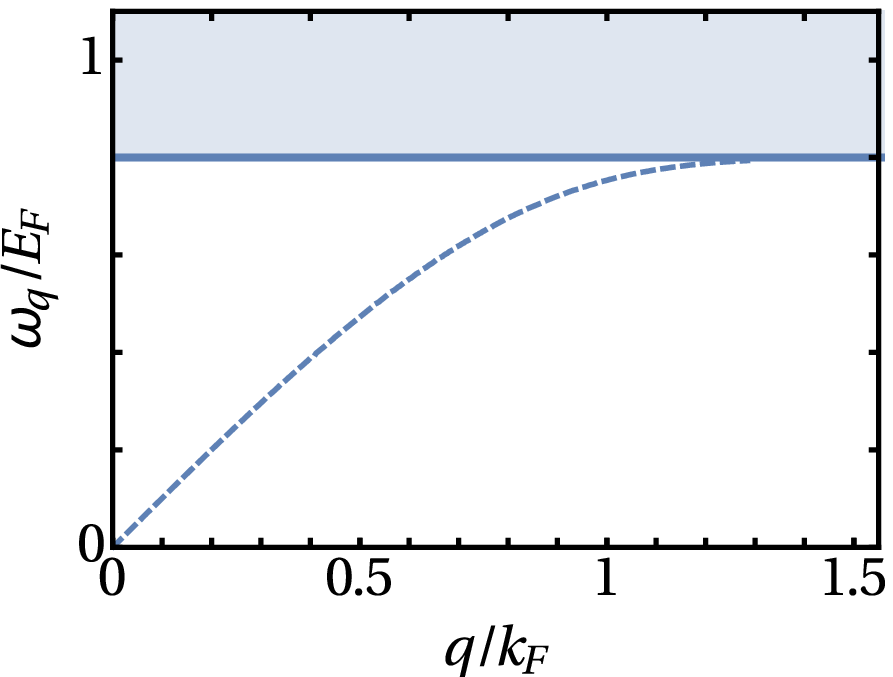}
          \end{figure}
      \end{minipage}
            \begin{minipage}{0.3\linewidth}\centering
          \flushleft (c) \vspace{-5mm} \begin{figure}[H]\centering
              \includegraphics[width=0.9\linewidth]{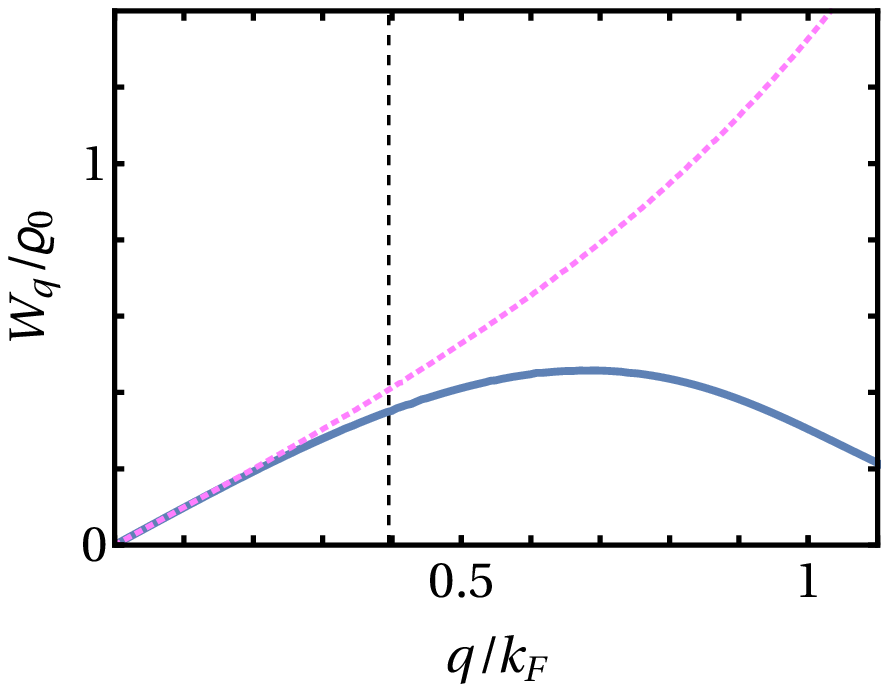}
          \end{figure}
      \end{minipage}
  \end{minipage}
  
    \begin{minipage}{\linewidth}\centering
     \begin{minipage}{0.3\linewidth}\centering
          \flushleft (d) \vspace{-8mm} \begin{figure}[H]\centering\hspace{-4mm}
			\includegraphics[width=0.95\linewidth]{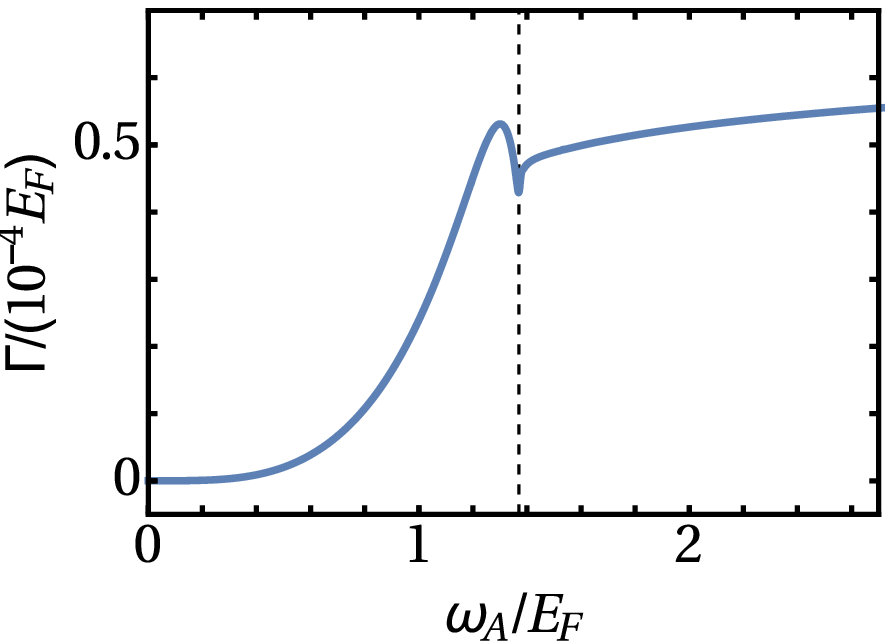}
          \end{figure}
      \end{minipage}
      \begin{minipage}{0.3\linewidth}\centering
          \flushleft (e) \vspace{-8mm} \begin{figure}[H]\centering
              \includegraphics[width=0.9\linewidth]{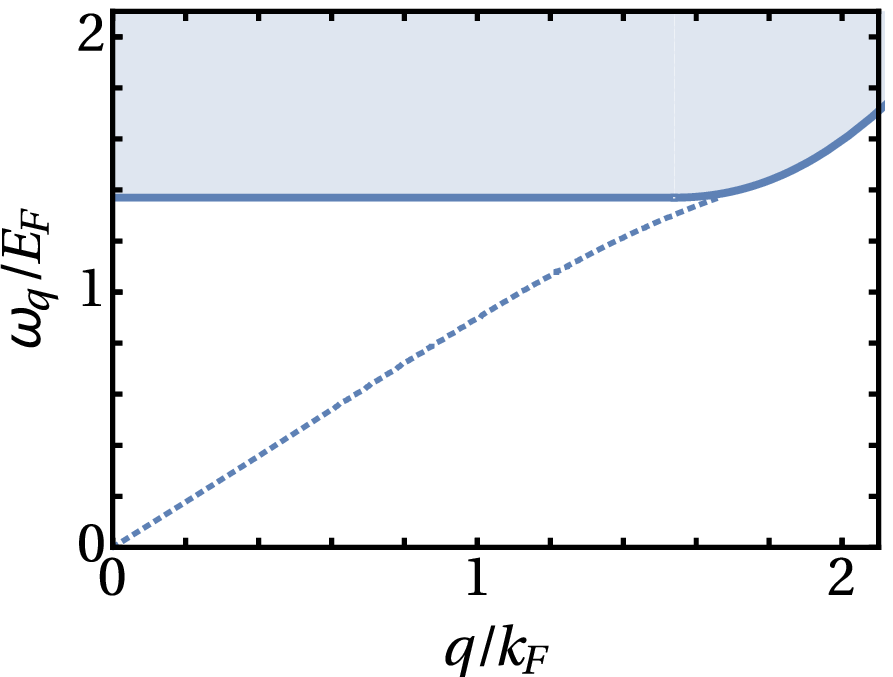}
          \end{figure}
      \end{minipage}
         \begin{minipage}{0.3\linewidth}\centering
          \flushleft (f) \vspace{-5mm} \begin{figure}[H]\centering
              \includegraphics[width=0.9\linewidth]{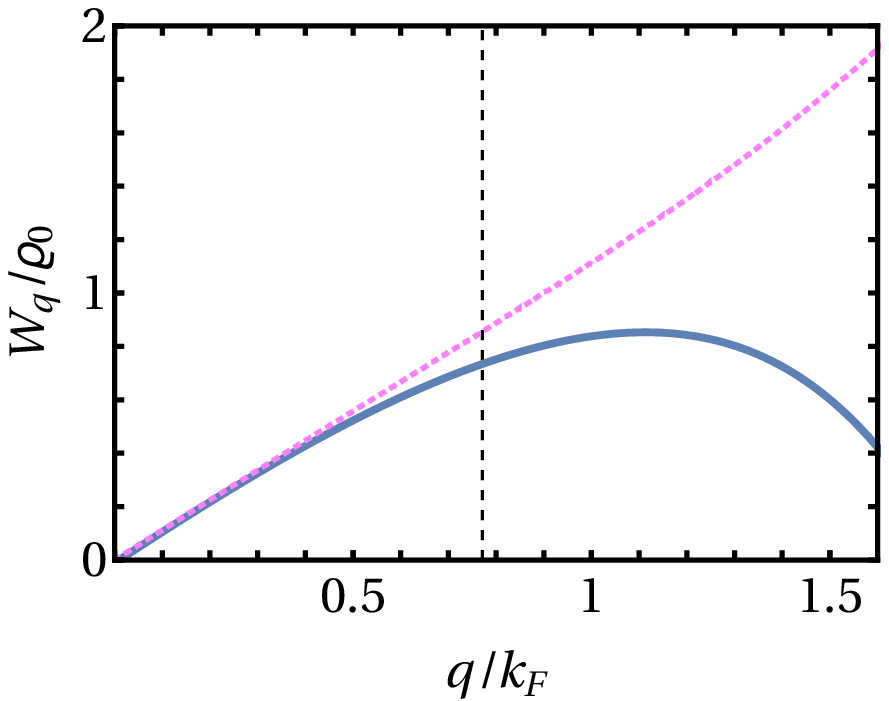}
          \end{figure}
      \end{minipage}
   \end{minipage}
     
          \begin{minipage}{\linewidth}\centering
     \begin{minipage}{0.3\linewidth}\centering
          \flushleft (g) \vspace{-8mm} \begin{figure}[H]\centering
			\includegraphics[width=0.9\linewidth]{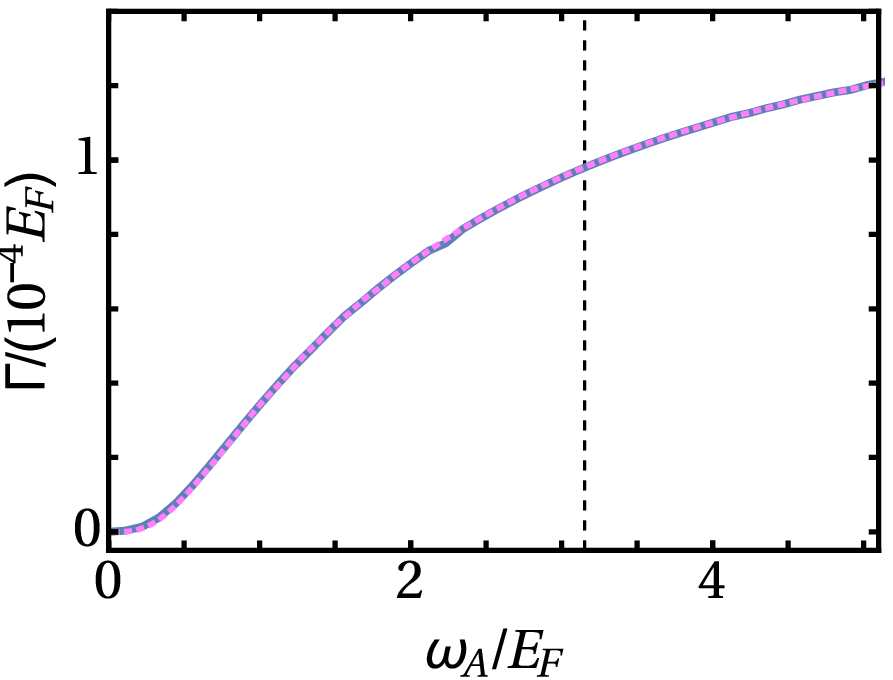}
          \end{figure}
      \end{minipage}
      \begin{minipage}{0.3\linewidth}\centering
          \flushleft (h) \vspace{-8mm} \begin{figure}[H]\centering
              \includegraphics[width=0.9\linewidth]{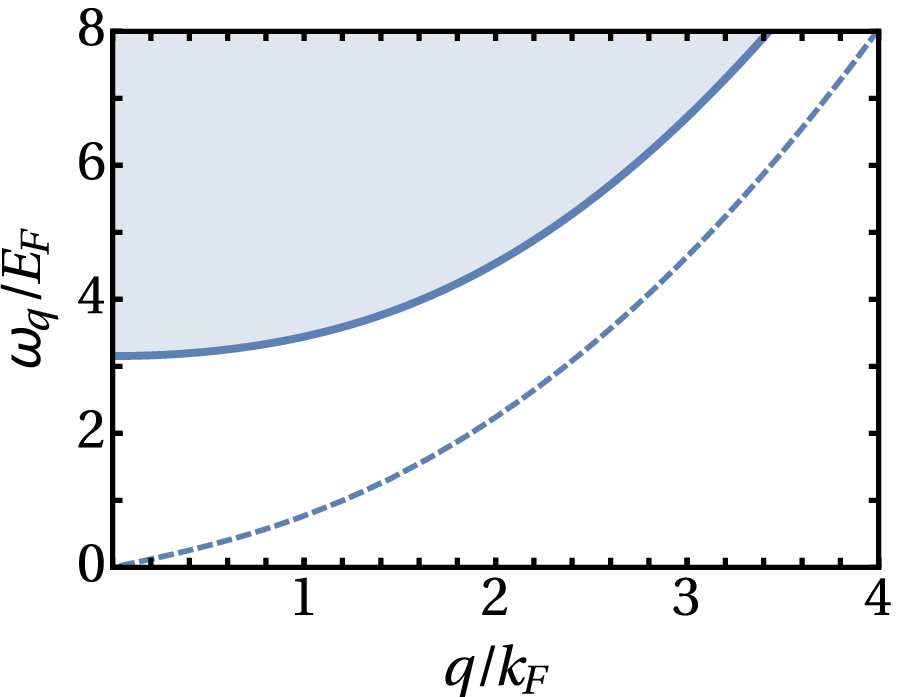}
          \end{figure}
      \end{minipage}
         \begin{minipage}{0.3\linewidth}\centering
          \flushleft (i) \vspace{-5mm} \begin{figure}[H]\centering
              \includegraphics[width=0.9\linewidth]{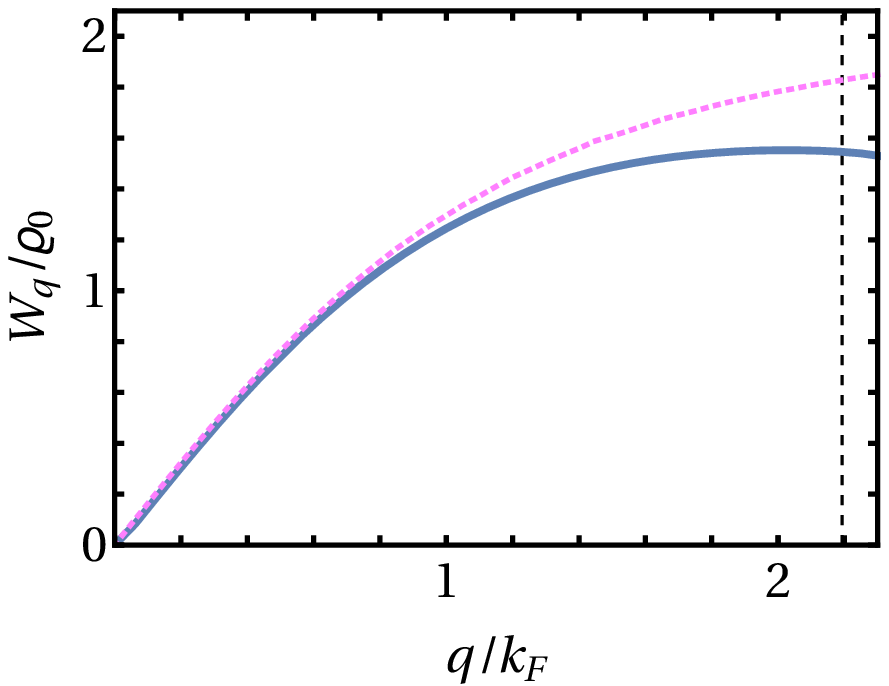}
          \end{figure}
      \end{minipage}
         \end{minipage}

\caption{Results at $T=0$ for (a-c) $1/k_F a_s = -0.5$, (d-f) $1/k_Fa_s = 0$, and (g-i) $1/k_Fa_s = 1.0$. (a,d,g) Impurity decay rate $\Gamma$ against vibrational frequency $\omega_A$ (solid line), with pair gap $\Theta_0$ (vertical dashed line); (g) contribution from the collective mode resonance only, i.e.\ neglecting Beliaev processes (pink dotted line). (b,e,h) Dispersion relation of the collective mode $\omega_q$ (dashed line) relative to the threshold frequency $\Theta_q$ (solid line) lower-bounding the pair continuum (shaded region), following Ref.~\cite{Combescot2006pra}. (c,f,i) Spectral weight $W_q$ of the collective mode (solid line), the long-wavelength approximation $W_q \approx \varrho_0 \varepsilon_q/\omega_q$ (dotted line), and the inverse coherence length $\zeta^{-1}$ (vertical dashed line). Parameters are $M/m = 40/6$ and $\kappa = 0.18 E_F/k_F^3$, which describes a $^{40}$K impurity in a $^{6}$Li gas at density $\varrho_0 = 2.1\times 10^{12}~$cm$^{-3}$, corresponding to $E_F \approx 2\pi\times 13~$kHz and $k_F\approx 2\pi/(160~$nm$)$ \cite{Spiegelhader2009}.\label{bcsPlots}}
\end{figure*}

We note in brief that a second decay channel for phonons exists at finite temperature due to the possibility of scattering from thermally excited fermionic quasiparticles, depicted in Fig.~\ref{BeliaevLandau}(b). This so-called Landau damping lead to a finite lifetime even for low-frequency phonons with $\omega_q \ll \Theta_0$. However, this lifetime can be assumed to be effectively infinite in the temperature regime of interest to us, $\beta\Theta_0\gg 1$, since the population of thermally excited quasiparticle pairs is negligible.

\section{Numerical results}
\label{sec_results}

We now illustrate our results by explicitly calculating the decay rate of the impurity for several examples. We take parameters from the experiments reported in Ref.~\cite{Spiegelhader2009}. The DSF is computed using Eq.~\eqref{fluctuationDissipationTheorem}, with $\epsilon$ finite but chosen small enough to obtain convergence; we have found $\epsilon = 0.01E_F$ to be adequate. We assume zero temperature, which yields a good approximation to the finite-temperature decay rate for the vibrational frequencies of greatest interest, $\omega_A \gtrsim \Theta_0$, given that we assume $\beta\Theta_0 \gg 1$. Note that the impurity confinement length $\ell=1/\sqrt{M\omega_A}$ entering the form factor \eqref{formFactor} also varies as a function of $\omega_A$ in our calculations. 

Our numerical results are summarized in Figs.~\ref{bcsPlots} and \ref{transitionPlots}, and described in detail in the following subsections.

\subsection{Measuring the pair gap}

\begin{figure*}
\centering
\begin{minipage}{\linewidth}\centering

\begin{minipage}{0.45\linewidth}\centering
\flushleft\hspace{1cm}(a)\vspace{-7mm}
\begin{figure}[H]\centering
\includegraphics[width=0.7\linewidth]{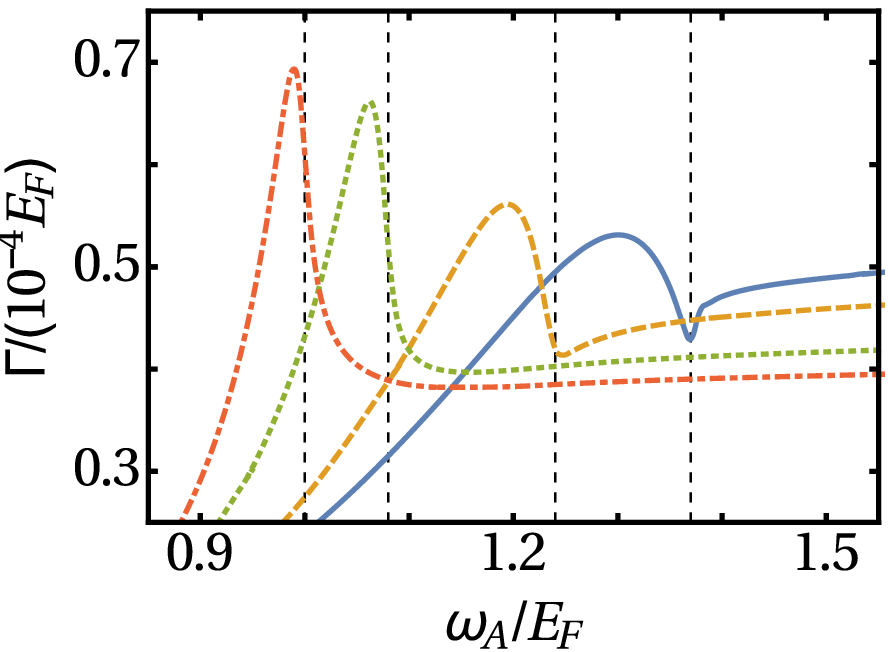}
\end{figure}
\end{minipage}
\begin{minipage}{0.45\linewidth}
\flushleft\hspace{1cm}(b)\vspace{-7mm}
\begin{figure}[H]\centering
\includegraphics[width=0.7\linewidth]{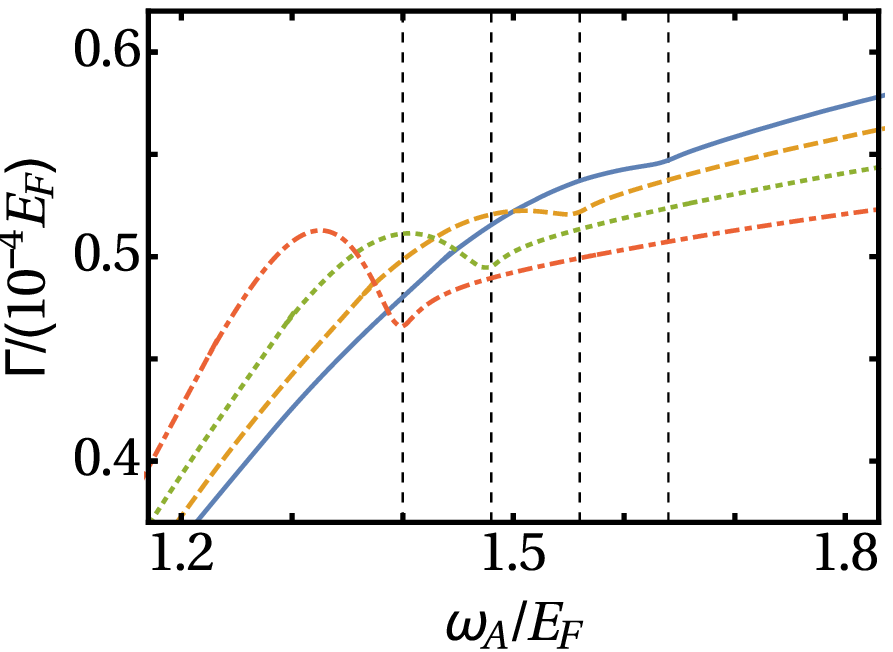}
\end{figure}
\end{minipage}
\end{minipage}
\caption{(a) Transition between BCS and unitary regimes, showing the decay rate for $1/k_Fa_s = -0.31$ (dot-dashed line), $-0.24$ (dotted line), $-0.11$ (dashed line), and $0$ (solid line). (b) Transition between unitary and BEC regimes, showing the decay rate for $1/k_Fa_s = 0.02$ (dot-dashed line), $0.08$ (dotted line), $0.14$ (dashed line), and $0.20$ (solid line). (a,b) The pair gap $\Theta_0$ for each example is shown by the vertical dashed lines. The parameters are the same as in Fig.~\ref{bcsPlots}. \label{transitionPlots}}
\end{figure*}

In the BCS and unitary regimes, the pair gap $\Theta_0$, and thus the absolute value of the order parameter $\Delta = \Theta_0/2$, can be inferred from the extremal values of the decay rate as a function of the impurity trapping frequency. The nature of this extremal behavior depends on the regime considered, as we now discuss.

In the BCS limit, with $1/k_F a_s$ large and negative, there is a sharp peak in the decay rate occuring exactly at $\omega_A = \Theta_0$.  This behavior extends even to the regime of fairly small, negative $1/k_F a_s$, as shown in Fig.~\ref{bcsPlots}(a) where we plot an example for $1/k_Fa_s = -0.5$. The peak at $\omega_A = \Theta_0$ is a consequence of the well-known divergence in the density of states in the BCS limit. To see this clearly, in Fig.~\ref{bcsPlots}(b) we plot the dispersion relation of the collective mode, which reproduces the calculation of Combescot et al.~\cite{Combescot2006pra}. The plot shows that $\omega_q$ bends to become almost flat as it approaches and then merges with the pair continuum. This indicates that the density of states $D(\nu) \sim \lvert \dd\omega_q/\dd q\rvert^{-1}_{\omega_q=\nu}$ increases sharply as $\nu \to \Theta_0$. 	

On the other hand, at unitarity quite a different behavior is found, as shown in Fig.~\ref{bcsPlots}(d). Here, the decay rate is a local minimum at $\omega_A = \Theta_0$. The collective mode dispersion $\omega_q$ remains approximately linear almost all the way into the continuum \cite{Combescot2006pra}, as plotted in Fig.~\ref{bcsPlots}(e). This implies that the phonon density of states grows approximately quadratically, $D(\nu) \sim \nu^2$, as the pair gap is approached. However, the increase of $D(\nu)$ is counteracted by the behavior of $W_q$ shown in Fig.~\ref{bcsPlots}(f), which decreases quite rapidly as the collective mode merges into the continuum. This property of the unitary gas can be interpreted as a ``stiffness'' of the ground state against excitation by single-particle scattering processes at frequencies comparable to the pair gap.  For higher frequencies $\omega_A>\Theta_0$, the decay rate increases abruptly due to the contribution of pair excitations, leading to a larger volume of momentum space available for scattering.

Note that the transition between these two qualitatively different behaviors occurs smoothly as unitarity is approached, as shown in Fig.~\ref{transitionPlots}(a). As a result, we find an intermediate regime $-0.3 \lesssim 1/k_Fa_s \lesssim -0.1$ in which the extremal points of $\Gamma$ do not coincide with the pair gap. It should be noted that our results depend on the BCS mean-field approximation, which is not expected to accurately describe the Fermi gas near unitarity. This regime would therefore be particularly interesting to explore experimentally.

\subsection{Separation of discrete and continuous spectra}
\label{sec_separationOfSpectra}

In the BEC limit, the decay rate is instead a monotonically increasing function of frequency, as shown in Fig.~\ref{bcsPlots}(g) for $1/k_Fa_s = 1.0$. In particular, there are no remarkable features at the pair gap $\omega_A = \Theta_0$. This can be understood from the fact that the collective mode no longer merges with the pair continuum for $1/k_Fa_s > 0.16$ \cite{Combescot2006pra}, as can be seen from Fig.~\ref{bcsPlots}(h). Mathematically, this means that the equation $\omega_\qq = E_{\kk+\qq/2} + E_{\kk-\qq/2}$ has no solution. Physically, this describes the impossibility of a phonon decaying into a quasiparticle pair while conserving energy and momentum. As a result, broadening of the phonon frequencies due to Beliaev damping is essentially negligible. 

In order to illustrate this, we have calculated the impurity decay rate assuming that only a single frequency $\nu=\omega_q$ contributes to $S(q,\nu)$ for each value of $\qq$, i.e.\ assuming that Eq.~\eqref{DSFcolective} holds even for $\nu > \Theta_0$. The result of this approximation is shown by the dotted line in Fig.~\ref{bcsPlots}(g), which coincides almost perfectly with the more accurate calculation taking Beliaev processes into account. Note that this implies that the direct production of quasiparticle pairs from the static condensate is also negligible. Therefore, the behavior at $1/k_Fa_s = 1.0$ already agrees with the standard picture of a Bose-Einstein condensate in which the only significant contribution to the density fluctuations comes from long-lived excitations of the collective mode~\cite{PinesNozieres}.

We have also calculated some examples from the transition regime $0< 1/k_F a_s \lesssim 0.2$, shown in Fig.~\ref{transitionPlots}(b). The feature at $\omega_A = \Theta_0$ is seen to vanish for rather small values of $1/k_Fa_s$, as the collective mode separates from the pair continuum.  By measuring the frequency dependence of $\Gamma$ for various values of $1/k_F a_s$, it should thus be possible to pinpoint the separation of discrete and continuous spectra in an experiment. 

\section{Conclusion}
\label{sec_conclusion}

We have studied the evolution of a qubit comprising two vibrational energy levels of an anharmonically confined impurity interacting with a neutral Fermi superfluid, as could be realized in a state-of-the-art cold-atom implementation. We have related the impurity's spectral density to the dynamic structure factor of the superfluid. Our results indicate that various features of the superfluid's energy spectrum along the BCS-BEC crossover can be probed using such an impurity. These include the divergence in the density of BCS quasiparticle states, the energy gap for pair excitations in the BCS and unitary regimes, and the separation of discrete and continuous spectra as the BEC limit is approached. These capabilities would complement other proposed schemes to optically detect and probe fermionic superfluidity \cite{Ruostekoski1999pra,Torma2000prl,Kinnunen2004prl,Challis2007prl}. Moreover, our setup provides a testbed for more sophisticated theories of the unitary Fermi gas that go beyond the simple BCS mean-field approximation used here.

Using experimentally realistic parameters, we predict decay rates on the order of $\Gamma\sim 10^{-4} E_F \sim 10~$Hz, which are measurable within the typical lifetime of a cold atomic gas, and could be further increased at higher superfluid densities. The examples that we have considered require impurities confined on length scales on the order of $\ell \gtrsim 0.1 k_F^{-1} \sim 10~$nm, which is straightforwardly achievable using a heavy impurity in an optical dipole trap. The initial state of the probe can be prepared using a moving optical lattice potential to excite the impurity's motion, while readout can be achieved by, for example, band mapping combined with time-of-flight imaging \cite{Bloch2008rmp}. In addition, we have shown that it is possible to greatly enhance the measurement signal by immersing many impurities within a single realization of the gas, in such a way that these individual probes remain independent from one another. We thus conclude that our proposal represents a feasible alternative to optical measurements that is nondestructive and offers nanometre-scale spatial resolution in principle. 

For simplicity, we have assumed that the superfluid is invariant under translations. A homogeneous Fermi superfluid can be realized in a uniform trap potential \cite{Mukherjee2016arx}, but our results are also relevant for harmonically trapped gases, so long as the impurities are situated far from the edge of the atomic cloud. Although long-wavelength collective modes are strongly modified by the presence of a harmonic trap \cite{Bruun2001prl}, we expect the high-frequency excitations with wavelengths much smaller than the Thomas-Fermi radius of the gas to remain essentially unaffected. Nevertheless, the inhomogeneous density of the harmonically trapped system leads to a spatially varying order parameter $\Delta(\rr)$ and chemical potential $\mu(\rr)$ \cite{Giorgini2008rmp}. As a result, averaging the response from multiple probes that are widely dispersed within a single superfluid sample would broaden the sharp features visible in Fig.~\ref{bcsPlots}(a,d), since each impurity would see a different local pair excitation gap $\Theta_0(\rr)$. Information on the spatial profile of the order parameter could be extracted by measuring the extent of this broadening in an inhomogeneous system. On the other hand, excessive broadening would eventually make it difficult to distinguish the features of interest, which restricts the size of the region within the gas cloud that could be probed --- and therefore the number of impurities that could be simultaneously used ---  in a single measurement.

Several promising avenues suggest themselves for further research on topics related to this work. For instance, it would be straightforward to generalize our model in order to describe transitions between multiple vibrational levels of the impurity. One interesting possibility to consider in this regard is the sympathetic cooling of harmonically trapped impurities using the Fermi superfluid as a cold reservoir \cite{Daley2004pra, Griessner2006prl, Griessner2007njp}. Our results already indicate that the cooling rate could be greatly increased in the BCS regime by tuning the impurity's trap frequency to equal the pair excitation gap  in order to take advantage of the divergent density of states. Another possible extension concerns the use of an impurity to probe exotic superfluid states that could arise, for example, in low-dimensional systems \cite{Liao2010nat,Mitra2016prl} or in the presence of synthetic gauge fields~\cite{Vyasankere2011prb,Wang2012prl,Cheuk2012prl}.

Beyond its potential for measuring properties of the gas itself, our setup constitutes an interesting open quantum system in its own right. An unusual feature of our model is the possibility of exciting both discrete and continuous quasiparticle modes in the reservoir, which moreover are coupled together nontrivially. This leads to distinctive features in the spectral density in the BCS and unitary regimes. In the BCS limit, the peak at frequencies commensurate with the pair gap is reminiscent of resonances appearing in the spectral density describing the vibrational environment of certain photosynthetic systems, which are currently the subject of intense scrutiny in the field of quantum biology \cite{Prior2010prl,Chin2013nat,Huelga2013cp}. Such peaks are typically associated with non-Markovian dynamics, although more sophisticated theoretical techniques would be required to explore this possibility in our case. Furthermore, at temperatures comparable to the pair gap we expect qualitatively new features to appear in the spectral density, due either to Landau damping or to the onset of the normal state as the critical condensation temperature is approached. Both of these effects could be explored within the present framework. 

In summary, a trapped impurity atom coupled to a neutral Fermi superfluid may constitute a useful experimental probe of density fluctuations, but also represents a novel platform to explore the physics of open quantum systems. The problem thus merits further theoretical and experimental study.

\begin{acknowledgments}
MTM acknowledges moral support from the Controlled Quantum Dynamics CDT, which is funded by EPSRC. DJ acknowledges EU support through project QuProCS Grant Agreement No.\ 641277.
\end{acknowledgments}


\input{DSFprobePaper.bbl}

\appendix

\input{masterEquation}
\input{DSFderivation}

\end{document}

%% file: packages.tex
\usepackage[T1]{fontenc}
\usepackage{graphicx,epsfig}
\usepackage{amsmath,amssymb,bbm,bm}
\usepackage{float}
\usepackage{fancyhdr}
\usepackage{import}
\usepackage{booktabs}
\usepackage{exscale}
\usepackage[english]{babel}
\usepackage{color}
\usepackage{appendix}
\usepackage{dsfont}
\usepackage{hyperref}
\usepackage{xcolor}
\hypersetup{
    colorlinks,
    linkcolor={rgb:red,2; blue,1},
    citecolor={blue!60!black},
    urlcolor={blue!60!black}
}

%% file: definitions.tex
\renewcommand{\d}{\dagger}

\newcommand{\ii}{\ensuremath{\mathrm{i}}}
\newcommand{\ee}{\ensuremath{\mathrm{e}}}
\newcommand{\dd}{\mathrm{d}}
\newcommand{\dD}[1]{\mathrm{d}^{#1}}

\newcommand{\dt}[1]{\ensuremath{\frac{\mathrm{d} #1}{\mathrm{d} t}}}

\newcommand{\bra}[1]{\ensuremath{\left\langle #1 \right\rvert}}
\newcommand{\ket}[1]{\ensuremath{\left\lvert #1 \right\rangle}}

\newcommand{\ketbra}[2]{\ensuremath{\left| #1 \middle\rangle\middle\langle #2\right|} }
\newcommand{\ketbrasub}[3]{\ensuremath{\left| #1 \middle\rangle_{#2}\middle\langle #3\right|} }
\newcommand{\braket}[2]{\ensuremath{\left\langle #1 \vphantom{#2} \middle\vert  #2 \vphantom{#1}\right\rangle}}
\newcommand{\matrixel}[3]{\ensuremath{\left\langle #1 \vphantom{#2#3} \middle\vert #2 \middle| #3 \vphantom{#1#2} \right\rangle}}

\renewcommand{\Im}{\mathrm{Im}}

\newcommand{\Tr}{\mathrm{Tr}}

\newcommand{\sinc}{\ensuremath{\mathrm{sinc}}}

\newcommand{\kk}{\ensuremath{\mathbf{k}}}

\newcommand{\qq}{\ensuremath{\mathbf{q}}}

\newcommand{\xx}{\ensuremath{\mathbf{x}}}

\newcommand{\rr}{\ensuremath{\mathbf{r}}}

\newcommand{\ZZ}{\ensuremath{\mathcal{Z}}}

\newcommand{\SJ}{\mathcal{J}}
\newcommand{\SI}{\mathcal{I}}



%% file: DSFprobePaper.bbl
%

%% file: masterEquation.tex
\newpage 
\section{Derivation of the master equation}
\label{app_superfluidME}

In this appendix, we provide technical details of the master equation and its derivation.

\subsection{Single impurity}
\label{sec_singleImpurityFermiSuperfluidAppendix}

We first derive the master equation describing a single impurity immersed in a neutral Fermi superfluid. In particular, we prove the assertion made in the main text that bath-induced dephasing and transitions between degenerate sublevels can be neglected. We consider an isotropic potential confining the impurity, therefore each energy level above the ground state is triply degenerate. Including only the lowest two energy levels, the autonomous Hamiltonian of the impurity is
\begin{equation}
\label{impurityDegenerateHa}
H_A = \sum_{a=1}^3 \omega_A\ketbra{a}{a},
\end{equation}
where $\ket{a}$ describes a motional excitation in the $x_a$ direction, with $\{x_1,x_2,x_3\} = \{x,y,z\}$, while the ground state is denoted $\ket{0}$. All higher motional states are assumed to be off-resonant. In the following, Latin indices such as $a = 1,2,3$ enumerate only states in the excited manifold, while Greek indices such as $\gamma = 0,1,2,3$ refer to all four impurity states.

The impurity-gas interaction is given by 
\begin{equation}
\label{multiStateImpurityGasInteraction}
H_{AB} = \sum_{\gamma,\delta= 0}^3 A_{\gamma\delta}  B_{\gamma\delta},
\end{equation}
where $A_{\gamma\delta} = \ketbra{\gamma}{\delta}$ and
\begin{equation}
\label{Bmn}
B_{\gamma\delta} = \frac{1}{V}\sum_{\qq\neq 0} \lambda_{\qq}^{(\gamma\delta)} \varrho_\qq,
\end{equation}
with 
\begin{equation}
\label{lambdaMNappendix}
\lambda^{(\gamma\delta)}_\qq = \kappa \int\dD{3}\rr \; \ee^{\ii\qq\cdot \rr} \phi_\gamma(\rr)\phi_\delta(\rr).
\end{equation}
The wave functions in the harmonic approximation are given by $\phi_0(\rr) \propto \ee^{-r^2/2\ell^2}$ and $\phi_a(\rr) \propto  x_a \ee^{-r^2/2\ell^2}$. We have neglected to write the normalization factors, which are chosen to be real numbers for simplicity. The coupling constants are found to be, for $a \neq b$,
\begin{align}
\label{lambdaMNfull}
\lambda_\qq^{(00)} & = \kappa \ee^{-\ell^2 q^2/4}, \notag \\
\lambda^{(a0)}_\qq & = \frac{\ii\kappa}{\sqrt{2}} \ell q_a \ee^{-\ell^2 q^2/4}, \notag \\
\lambda^{(ab)}_\qq & = -\frac{\kappa}{2} \ell^2 q_a q_b \ee^{-\ell^2 q^2/4}, \notag \\
\lambda^{(aa)}_\qq & = \kappa \left (1 - \frac{\ell^2 q^2}{2} \right ) \ee^{-\ell^2 q^2/4},
\end{align}
where $\{q_1,q_2,q_3\} =  \{q_x,q_y,q_z\}$ denote the Cartesian components of $\qq$. Our choice of real-valued wave functions means that $\lambda^{(\delta\gamma)}_\qq = \lambda_\qq^{(\gamma\delta)}$.

We now derive a master equation for the impurity density operator under the Born-Markov and rotating-wave approximations, assuming an initial product state of the form $\rho(0) = \rho_A(0)\rho_B$ with $\rho_B = \ee^{-\beta H_B}/\ZZ_B$, as described in Section~\ref{sec_fermiME} (see, for example, Section 3.3 of Ref.~\cite{BreuerPetruccione} for details). The result takes the form
\begin{widetext}
\begin{equation}
\label{SuperfluidMEfull}
\dt{\rho_A} = \ii[\rho_A, H_A] + \ii\sum_{\gamma,\delta} \Xi_{\gamma\delta} [\rho_A,A_{\gamma\delta}] + \sum_{\gamma,\delta,\eta,\zeta} \Gamma_{\gamma\delta\eta\zeta} \left ( A_{\gamma\delta} \rho_A A_{\eta\zeta}^\d - \frac{1}{2} \{ A_{\eta\zeta}^\d A_{\gamma\delta},\rho_A\} \right ).
\end{equation}
\end{widetext}
Here, we introduced the incoherent rates 
\begin{equation}
\label{superfluidGammaFull}
\Gamma_{\gamma\delta\eta\zeta}  = 2\pi \SI_{\gamma\delta\eta\zeta}(\omega_{\delta\gamma}),\\
\end{equation}
where $\omega_{\delta\gamma} = \omega_\delta - \omega_\gamma$ is the difference in energy between states $\ket{\gamma}$ and $\ket{\delta}$, i.e.\
\begin{equation}
\label{wnmDef}
\omega_{\delta\gamma} = \left \lbrace 
\begin{array}{ll}
\omega_A& \quad (\gamma=0,\delta\neq 0)\\
-\omega_A& \quad (\delta=0,\gamma\neq 0)\\
0	& \quad (\mathrm{otherwise}),
\end{array}\right. 
\end{equation}
and the spectral densities are defined by
\begin{equation}
\label{ImnSuperfluidDef}
\SI_{\gamma\delta\eta\zeta}(\nu) = \frac{1}{V}\sum_{\qq\neq 0}  \left (\lambda_\qq^{(\gamma\delta)}\right )^* \lambda^{(\eta\zeta)}_\qq S(\qq,\nu),
\end{equation}
valid for $\omega_{\delta\gamma} = \omega_{\zeta\eta}$, while $\SI_{\gamma\delta\eta\zeta}(\nu) = 0$ otherwise. We also defined the energy shifts
\begin{equation}
\label{superfluidXiFull}
\Xi_{\gamma\delta} = \sum_{\eta} \mathsf{P}\int_0^\infty\dd\nu\; \frac{\SI_{\eta\delta \eta\gamma}(\nu)}{\omega_{\delta\eta} - \nu},
\end{equation}
where $\mathsf{P}$ denotes the principal value.

For completeness, we briefly recap the steps involved in the derivation of the master equation \eqref{SuperfluidMEfull}. Working in an interaction picture with respect to the free Hamiltonian $H_A+H_B$ and at second order in the perturbation $H_{AB}$, the dynamics of the impurity is determined by the reservoir correlation functions 
\begin{equation}
\label{corrFunctionMEDerivation}
G_{\gamma\delta\eta\zeta}(t) = \int_0^t\dd t'\int_{-\infty}^\infty\dd \nu\; \ee^{\ii (\omega_{\delta\gamma}-\nu) t'} \SI_{\gamma\delta\eta\zeta}(\nu).
\end{equation}
It is assumed that each of these correlation functions converges quickly to a stationary value. This convergence occurs after the reservoir correlation time $\tau_B$, which must be much smaller than any relevant time scale of the impurity's evolution in the interaction picture. This assumption justifies the Markov approximation, in which $G_{\gamma\delta\eta\zeta}(t)$ is replaced by its asymptotic value as $t\to \infty$, whose real and imaginary parts give rise to the incoherent rates $\Gamma_{\gamma\delta\eta\zeta}$ and energy shifts $\Xi_{\gamma\delta}$, respectively. Self-consistency of the Markov approximation requires that each spectral density $\SI_{\gamma\delta\eta\zeta}(\nu)$ must be approximately constant over frequency changes of order $\Gamma_{\gamma\delta\eta\zeta}$ or $\Xi_{\gamma\delta}$ around the central frequency $\omega_{\delta\gamma}$ of the reservoir-induced transition in question. 

After the Markov approximation has been made, the master equation contains counter-rotating terms that oscillate at frequency $\pm 2\omega_A$ and are proportional to the rates $\Gamma_{\gamma\delta\eta\zeta}$ and $\Xi_{\gamma\delta}$. The effect of these terms averages to zero over the time scale relevant for the impurity's evolution, assuming that $\omega_A \gg \Gamma_{\gamma\delta\eta\zeta},|\Xi_{\gamma\delta}|$. If this condition holds, the counter-rotating terms can be neglected (rotating-wave approximation), whence we obtain the Lindblad equation~\eqref{SuperfluidMEfull}.

At this stage we make use of the inversion symmetry of the superfluid, which implies that $S(\qq,\nu) = S(-\qq,\nu)$. This leads to an enormous simplification, since most of the $256$ functions $\SI_{\gamma\delta\eta\zeta}(\nu)$ can be shown to vanish identically. In particular, $\SI_{\gamma\delta\eta\zeta}(\nu) = 0$ if any pair of indices are equal to each other while the other pair of indices are not equal to each other. Furthermore, $\SI_{\gamma\delta\eta\zeta}(\nu) = 0$ if all four indices are different. This leaves in total $40$ nonzero functions of the form $\SI_{\gamma \delta \gamma \delta}(\nu) = \SI_{\delta\gamma\gamma\delta}(\nu)$ or $\SI_{\gamma\gamma\delta\delta}(\nu)$. However, due to isotropy [$S(\qq,\nu) = S(q,\nu)$] and the index permutation symmetry [$\lambda^{(\delta\gamma)}_\qq = \lambda_\qq^{(\gamma\delta)}$] we are actually left with only five independent functions: $\SI_{0000}(\nu)$, $\SI_{00aa}(\nu)$, $\SI_{aabb}(\nu)$, $\SI_{0a0a}(\nu)$, and $\SI_{abab}(\nu)$, for $a\neq b$. 

Referring to Eq.~\eqref{superfluidXiFull}, the aforementioned conditions imply that $\Xi_{\gamma\delta}$ vanishes unless $\gamma=\delta$. The nonzero terms $\Xi_{\gamma\gamma}$ describe a simple energy shift for each state. Furthermore, each sublevel of the excited state receives an identical shift relative to the ground state due to rotation symmetry. These energy shifts can be absorbed into the definition of the trapping frequency $\omega_A$ and shall be ignored from here on. 

Regarding the incoherent part of the master equation \eqref{SuperfluidMEfull}, three types of terms can be distinguished. First we consider dephasing processes, which are governed by terms in Eq.~\eqref{SuperfluidMEfull} with $\gamma=\delta$ and $\eta=\zeta$. The corresponding rates of the form $\Gamma_{\gamma\gamma\eta\eta}$ can be shown to vanish even at finite temperature, assuming that $\beta\Theta_0 \gg 1$. To demonstrate this, we use the fact that, as shown in Appendix~\ref{app_DSFderivation}, the dynamic structure structure factor in the limit $\nu \to 0$ is given by
\begin{equation}
\label{dephasingDSF}
S(q,\nu) = \frac{\varrho_0 q}{2m c} \left [ (1+n_q) \delta(\nu-\omega_q) + n_q \delta(\nu + \omega_q) \right ],
\end{equation}
valid for $\beta\Theta_0\gg 1$, where $n_q = (\ee^{\beta\omega_q} - 1)^{-1}$ and $\omega_q = cq$. Plugging this into the definitions \eqref{superfluidGammaFull} and \eqref{ImnSuperfluidDef}, we obtain, for example,
\begin{align}
\label{gammaDephasingSuperfluid}
\Gamma_{0000} & = \lim_{\nu \to 0} \frac{2\pi}{V} \sum_{\qq\neq 0} \lvert \lambda_\qq^{(00)}\rvert^2 S(\qq,\nu) \notag\\
& =  \frac{\kappa^2 \rho_0}{2\pi m c^5}\lim_{\nu \to 0}\left [ \nu^3 \coth\left (\frac{\beta \nu}{2}\right )\right ]\notag \\
 & = 0.
\end{align}
A similar argument demonstrates that $\Gamma_{00aa} = \Gamma_{aa00}= \Gamma_{aabb} = 0$, therefore dephasing processes do not contribute to the master equation. 

The second class of incoherent process that we consider comprises transitions between the three states in the excited manifold, occurring at a rate $\Gamma_{abab}$. It is straightforward to show that this rate also vanishes, following an analogous argument to the one presented in Eq.~\eqref{gammaDephasingSuperfluid}. In particular, we have that $\SI_{abab}(\nu)\sim \nu^6$ as $\nu\to 0$. This shows that the rate of incoherent transitions between these sublevels is negligible, even if an anisotropic perturbation breaks the degeneracy and a small energy difference $\nu$ exists between these states.

The third kind of dissipative process corresponds to incoherent transitions between the ground and excited states, occurring at the rates $\Gamma_{0a0a}$ and $\Gamma_{a0a0}$. If we assume that $\beta\omega_A\gg 1$, the detailed balance condition $S(\qq,-\nu) = \ee^{-\beta\nu}S(\qq,\nu)$ implies that the rates $\Gamma_{a0a0}$ are vanishingly small. Therefore, the probability of a transition from the ground state to one of the excited states is negligible. The state of the system thus remains within the subspace comprising the ground state and whichever excited motional state is addressed in the experiment. We conclude that the two-level approximation is valid, so long as $\beta\omega_A\gg 1$ and $\beta\Theta_0\gg 1$. Within the qubit subspace, the master equation \eqref{SuperfluidMEfull} is equivalent to Eq.~\eqref{fermiSuperfluidME} in the main text, with $\Gamma = \Gamma_{0a0a}$.

\subsection{Multiple impurities}

Now we consider the case where $N$ impurities are immersed in the superfluid. In this subsection, we explicitly model only two vibrational states for each impurity, which will be shown to be a self-consistent approximation. We assume that the energy splitting of each impurity qubit is identical, leading to the autonomous Hamiltonian
\begin{equation}
\label{multiImpuritySuperfluidHa}
H_A = \sum_{n=1}^N \omega_A\sigma^\d_n \sigma_n,
\end{equation}
where $\sigma_n = \ketbrasub{0}{n}{1}$ is the usual lowering operator pertaining to impurity $n$.

We assume that the minimum of the potential confining impurity $n$ is at position $\xx_n$. The excited state of qubit $n$ corresponds to motion in the direction parallel to the unit vector $\hat{\mathbf{d}}_n$, which may be different for each impurity. The vector $\hat{\mathbf{d}}_n$ can be understood as a kind of dimensionless acoustic dipole moment associated with each impurity. The interaction between the impurities will be seen to depend on the relative orientation of these dipole moments and their mutual separation vector. 

We write the interaction between the impurities and the Fermi gas as
\begin{align}
\label{multiImpurityFermiInteraction}
H_{AB} & = \kappa\sum_{n=1}^{N}\sum_{i,j=0}^1\ketbrasub{i}{n}{j} \int\dD{3}\rr\;\left (\phi_i^{(n)}(\rr)\right )^* \phi_j^{(n)}(\rr)\varrho(\rr) \notag \\
& = \frac{1}{V}\sum_{n=1}^{N}\sum_{i,j=0}^1\ketbrasub{i}{n}{j} \sum_{\qq\neq 0} \lambda_{\qq,n}^{(ij)} \ee^{\ii\qq\cdot \xx_n} \varrho_\qq,
\end{align}
where $\phi_i^{(n)}(\rr)$ denotes the wave function of impurity $n$ in state $\ket{i}$, and the coupling constants are
\begin{equation}
\label{lambdaQjFermi}
\lambda_{\qq,n}^{(ij)} = \kappa\int\dD{3}\rr\; \ee^{\ii\qq\cdot\rr} \left (\phi_i^{(n)}(\rr+\xx_n)\right )^* \phi_j^{(n)}(\rr+\xx_n).
\end{equation}
Upon deriving the master equation, the terms in Eq.~\eqref{multiImpurityFermiInteraction} with $i=j$ give rise to three contributions: dephasing, yet the rate for this vanishes according to the arguments in Section~\ref{sec_singleImpurityFermiSuperfluidAppendix}; local energy shifts, which can be absorbed into the definition of the impurity trap frequencies; and a bath-mediated interaction between the qubits of the form $\sigma^z_m\sigma^z_n$, where $\sigma^z_n = [\sigma^\d_n,\sigma_n]$, which does not affect the dynamics of populations in the energy eigenbasis. Since we are interested in the evolution of states that are diagonal in energy, we ignore these terms and, defining $\lambda_{\qq,n} = \lambda_{\qq,n}^{(10)}$, obtain the simplified interaction Hamiltonian
\begin{equation}
H_{AB} = \frac{1}{V}\sum_{n} \sum_{\qq} \left (\lambda_{\qq,n}\ee^{\ii\qq\cdot\xx_n} \sigma^\d_n\varrho_\qq + \lambda_{\qq,n}^* \ee^{-\ii\qq\cdot\xx_n}\sigma_n\varrho_\qq^\d \right ).
\end{equation}

We now derive a master equation describing the impurity density operator $\rho_A$ under the same approximations as in Appendix~\ref{sec_singleImpurityFermiSuperfluidAppendix}. Neglecting the bath-induced renormalization of the local qubit energies, and working at temperatures such that $\beta\omega_A\gg 1$, the master equation reads as
\begin{widetext}
\begin{equation}
\label{multiImpurityFermiME}
\dt{\rho_A} = \ii [\rho_A,H_A] + \ii \sum_{m\neq n} \eta_{mn} [\rho_A, \sigma^\d_m\sigma_n] + \sum_{m,n} \Gamma_{mn} \left ( \sigma_m \rho_A \sigma^\d_n - \frac{1}{2}\{\sigma^\d_n\sigma_m, \rho_A\} \right ).
\end{equation}
\end{widetext}
The couplings appearing above are given in terms of the spectral densities
\begin{equation}
\label{genSDjkFermi}
\SI_{mn}(\nu) = \frac{1}{V}\sum_{\qq}\lambda_{\qq,m}^*   \lambda_{\qq,n} \cos\left (\qq\cdot\xx_{mn}\right ) S(\qq,\nu),
\end{equation}
where $\xx_{mn} = \xx_m-\xx_n$ denotes the separation between impurities $m$ and $n$. Note that $\SI_{mn}(\nu)$ reduces to our standard definition \eqref{generalisedSpectralDensity} for $m=n$. The parameters entering the master equation are given explicitly by
\begin{align}
\label{GammaJKfermi}
\Gamma_{mn}& = 2\pi \SI_{mn}(\omega_A),\\
\label{etaJKfermi}
\eta_{mn} &= \mathsf{P}\int_0^\infty\dd\nu\; \frac{\SI_{mn}(\nu)}{\omega_A-\nu}.
\end{align}

Our goal now is to find a spatial configuration such that $\SI_{mn}(\nu) = 0$ for $m\neq n$, so that the impurities effectively decouple from one another. Evaluating the spectral densities in the thermodynamic limit, we use rotation invariance to obtain
\begin{equation}
\label{spectralDensityFormFactorMultiQubit}
\SI_{mn}(\nu) = \frac{ \kappa^2}{2\pi^2}\int_0^\infty\dd q \; q^2 \Phi_{mn}(q) S(q,\nu),
\end{equation}
where we defined the dimensionless form factors
\begin{equation}
\label{multiQUbitFOrmFactor}
\Phi_{mn}(q) = \frac{1}{4\pi\kappa^2} \int_{S^2} \dd\Sigma_\qq\; \lambda_{\qq,m}^* \lambda_{\qq,n} \cos(\qq\cdot \xx_{mn}).
\end{equation}
Using the isotropic, harmonic approximation for the impurity wave functions, we have that $\phi_i^{(n)}(\rr+\xx_n) \propto H_i(\hat{\mathbf{d}}_n\cdot \rr/\ell) \ee^{-r^2/2\ell^2}$, where $H_i(z)$ denotes the $i^{\mathrm{th}}$ Hermite polynomial for $i=0,1$. Hence, the leading-order contribution to Eq.~\eqref{multiQUbitFOrmFactor} as $x_{mn}\to \infty$ is 
\begin{align}
\label{farFieldFormFactor}
\Phi_{mn}(q) \approx & \,\frac{1}{2}\ell^2 q^2 \ee^{-\ell^2 q^2/2}\sinc(qx_{mn})\notag  \\ & \times \left ( \hat{\mathbf{d}}_m\cdot \hat{\xx}_{mn}\right )\left ( \hat{\mathbf{d}}_n\cdot \hat{\xx}_{mn}\right ),
\end{align}
where $\hat{\xx}_{mn}$ is a unit vector parallel to $\xx_{mn}$. This approximation is valid in the limit $q x_{mn}\gg 1$, and since only values $q \sim \ell^{-1}$ contribute to Eq.~\eqref{farFieldFormFactor} significantly, this corresponds to the far-field condition $x_{mn} \gg \ell$. 

Now, consider the case where the motion of each impurity is excited in the same direction $\hat{\mathbf{d}}_n = \hat{\mathbf{d}}$. By arranging the impurities in a regular array with lattice vector $\mathbf{b}$, such that $\mathbf{b}\cdot \hat{\mathbf{d}} = 0$ and $b \gg \ell$, it follows from Eq.~\eqref{farFieldFormFactor} that $\Phi_{mn}(q) \approx 0$ for $m\neq n$. In such a configuration, the impurities evolve independently according to the master equation \eqref{fermiSuperfluidME} in the main text. Note also that Eq.~\eqref{farFieldFormFactor} implies that the population of each impurity's motional states in directions orthogonal to $\hat{\mathbf{d}}$ are not influenced by the other impurities in the lattice. Therefore, the two-level approximation is self-consistent. 

%% file: DSFderivation.tex
\section{Dynamic structure factor}
\label{app_DSFderivation}

In this appendix, we give details of the dynamic structure factor $S(q,\nu)$, which is obtained via the fluctuation-dissipation theorem
\begin{equation}
\label{flucDissApp}
S(q,\nu) = \frac{-1}{\pi(1-\ee^{-\beta\nu})}\Im\left [\chi(q,\nu+\ii\epsilon)\right ],
\end{equation}
using the susceptibility $\chi(q,\nu)$ calculated in Refs.~\cite{Combescot2006pra,Guo2013ijmpb,Guo2013jltp,Cote1993prb,Zou2010pra,He2016ap}. We focus exclusively on the limit $\beta\Theta_0 \gg 1$, in which case the susceptibility simplifies considerably due to the absence of Landau processes of the type illustrated in Fig.~\ref{BeliaevLandau}(b). In order to present the result compactly, we use a shorthand notation where unprimed variables carry the subscript $\kk-\qq/2$, while primed variables carry the subscript $\kk+\qq/2$, e.g.\ $E = E_{\kk-\qq/2}$ and $E'=E_{\kk+\qq/2}$. The response function then takes the form
\begin{equation}
\label{chiSum}
\chi(q,\nu) = \chi_\mathrm{pair}(q,\nu) + \chi_\mathrm{coll}(q,\nu),
\end{equation}
where the pair contribution reads as
\begin{align}
\label{chiQP}
\chi_\mathrm{pair}(q,\nu) = -\frac{1}{V}\sum_\kk & \left \lbrace \frac{EE' - \xi\xi' + \Delta^2}{(E+E')^2 - \nu^2} \frac{E + E'}{EE'}  \right \rbrace.
\end{align}
The collective mode contribution is given by
\begin{equation}
\label{chiColl}
\chi_\mathrm{coll}(q,\nu) = \Delta^2\frac{A_1^2I_{11} + \nu^2 A_2^2 I_{22} - 2\nu^2 A_1 A_2 I_{12}}{I_{11}I_{22}-\nu^2 I^2_{12}},
\end{equation}
where we defined
\begin{align}
\label{A1}
A_1(q,\nu) & = \frac{1}{V}\sum_\kk  \frac{\xi + \xi'}{(E+E')^2 - \nu^2}  \frac{E + E'}{E E'}, \\
\label{A2}
A_2(q,\nu) & = \frac{1}{V}\sum_\kk  \frac{1}{(E+E')^2 - \nu^2} \frac{E+ E'}{EE'} , \\
\label{I11}
I_{11}(q,\nu) & = \frac{1}{V}\sum_\kk  \left ( \frac{EE' + \xi\xi' + \Delta^2 }{(E + E')^2 - \nu^2} \frac{E + E'}{EE'}  - \frac{1}{E_\kk}\right ), \\
\label{I22}
I_{22}	(q,\nu) & = \frac{1}{V}\sum_\kk  \left ( \frac{ EE' + \xi\xi' - \Delta^2 }{(E+E')^2 - \nu^2}  \frac{E + E' }{EE'} - \frac{1}{E_\kk} \right ), \\
\label{I12}
I_{12}(q,\nu) & = \frac{1}{V}\sum_\kk  \frac{1}{(E+E')^2 - \nu^2} \frac{E\xi' + E'\xi}{EE'}.
\end{align}

For frequencies $|\nu| < \Theta_0$, the imaginary parts of Eqs.~\eqref{chiQP} and \eqref{A1}--\eqref{I22} (evaluated at frequency $\nu + \ii\epsilon$) are zero. Therefore, the only contribution to the DSF comes from the pole of Eq.~\eqref{chiColl} corresponding to the collective mode resonance. To see this, we write
\begin{equation}
\label{chiCollCompact}
\chi_\mathrm{coll}(q,\nu) = \frac{B(q,\nu)}{\nu^2-\Omega^2(q,\nu)},
\end{equation}
where
\begin{align}
\label{Bqnu}
B(q,\nu) & = - \Delta^2 \frac{A_1^2I_{11} + \nu^2 A_2^2 I_{22} - 2\nu^2 A_1 A_2 I_{12}}{I_{12}^2}, \\
\label{OmegaQnu}
\Omega(q,\nu) & = \frac{\sqrt{I_{11}I_{22}}}{I_{12}}.
\end{align}
The dispersion relation of the collective mode is given by the solution of the nonlinear equation $\omega_q = \Omega(q,\omega_q)$. The spectral weight is defined as 
\begin{equation}
\label{MqDef}
W_q = \frac{B(q,\omega_q)}{2\omega_q}\left \lvert 1 - \frac{\dd \Omega(q,\nu)}{\dd\nu}\right \lvert^{-1}_{\nu=\omega_q}.
\end{equation}
Combining Eqs.~\eqref{flucDissApp} and \eqref{chiCollCompact} then leads directly to the DSF
\begin{equation}
\label{DSFcolectiveApp}
S(q,\nu) = W_q \left [ (1 + n_q)\delta(\nu - \omega_q) + n_q \delta(\nu + \omega_q) \right ],
\end{equation}
in agreement with Eq.~\eqref{DSFcolective}.

Note that the DSF at $T=0$ can be derived from Eq.~\eqref{chiCollCompact} in the alternative form 
\begin{equation}
\label{ScollQdelta}
S(q,\nu) = C_\nu \delta(q - q_\nu),
\end{equation}
where $\omega_{q_\nu} = \nu$ and we defined
\begin{equation}
\label{CqDirectDef}
C_\nu = \frac{B(q_\nu,\nu)}{2\nu}\left \lvert \frac{\dd \Omega(q,\nu)}{\dd\nu}\right \lvert^{-1}_{q=q_\nu}.
\end{equation}
However, Eq.~\eqref{ScollQdelta} can also be obtained directly from Eq.~\eqref{DSFcolectiveApp} by a simple change of variables in the delta function, from which it follows that
\begin{equation}
\label{CqDphRelation}
W_{q_\nu} D(\nu) = \frac{q_\nu^2}{2\pi^2} C_{\nu},
\end{equation}
where $D(\nu)$ is the phonon density of states defined in Eq.~\eqref{phononDensityOfStates}. Eqs.~\eqref{CqDirectDef} and \eqref{CqDphRelation} provide an efficient way to compute the spectral density \eqref{phononSpectralDensity} for $\nu < \Theta_0$. This method is used in particular to generate the pink dotted line in Fig.~\ref{bcsPlots}(g).


In the limit $q \to 0$ and $\nu\to 0$, we can find explicit analytical expressions for $\omega_q$ and $W_q$. To do so, we need the identity \cite{Combescot2006pra}
 \begin{equation}
 \label{massConservationIntegralIdentity}
I_{11}(q,\nu) = \frac{1}{V}\sum_{\kk}\frac{\nu^2 - (\kk\cdot\qq/m)^2}{(E+E')^2 - \nu^2} \frac{E + E'}{2E E'}.
 \end{equation}
This makes it straightforward to check, in the low-frequency and long-wavelength limit, that $I_{11}(q,\nu) \approx \nu^2 J_2/4 - q^2 J_4/12m^2$, as well as $A_1(0,0) = J_\xi$, $A_2(0,0) = J_2/2$, $I_{12}(0,0) = J_\xi/2$, and $I_{22}(0,0) = -\Delta^2 J_2$, where $J_2$, $J_4$ and $J_\xi$ are given in Eq.~\eqref{J2}. In this approximation, the solution of $\omega_q = \Omega(q,\omega_q)$ is found to be simply $\omega_q = cq$, with $c$ the sound speed given by Eq.~\eqref{soundSpeed}. The collective mode response function is approximately
 \begin{equation}
\label{chiLowQ}
\chi_\mathrm{coll}(q,\nu) \approx \frac{\Delta^2 J_2^2\nu^2 + J_\xi^2 \omega_q^2}{J_2(\nu^2 - \omega_q^2)}.
\end{equation}
Upon using the relation $\Delta^2 J_4 = 3m \varrho_0$ \cite{Combescot2006pra}, we find a DSF of the form Eq.~\eqref{DSFcolectiveApp} with $W_q = \varrho_0 \varepsilon_q/\omega_q$, which exhausts the $f$-sum and compressibility sum rules
\begin{align}
\label{fSumRule}
\int_{-\infty}^\infty \dd\nu\;\nu S(q,\nu) & = \varrho_0 \varepsilon_q, \\
\label{compressibilitySumRule}
\int_{-\infty}^\infty \dd\nu\;\lim_{q\to 0} \frac{S(q,\nu)}{\nu} & = \frac{\varrho_0}{2m c^2}.
\end{align}

At frequencies $|\nu|>\Theta_0$, the imaginary parts of $\chi_\mathrm{pair}(q,\nu+\ii\epsilon)$, $\Omega(q,\nu+\ii\epsilon)$ and $B(q,\nu+\ii\epsilon)$ are all nonzero, reflecting the contribution of fermionic pair excitations at frequencies above the gap. For example, direct pair production from the static condensate is described by the term
\begin{align}
\label{Spair}
S_\mathrm{pair}(q,\nu) & = -\frac{1}{\pi}\Im\left [ \chi_\mathrm{pair}(q,\nu+\ii\epsilon)\right ] \notag \\
&  = \frac{1}{V}\sum_\kk \frac{EE' - \xi\xi'+\Delta^2}{2EE'}\delta(\nu - E-E'),
\end{align}
assuming as usual that $\beta\Theta_0\gg 1$. The imaginary part of $\Omega(q,\nu+\ii\epsilon)$ can be interpreted as the damping rate of collective mode excitations due to Beliaev decay. This broadens the pole of Eq.~\eqref{chiColl}, ultimately leading to contributions to the DSF of Lorentzian-like form. However, the explicit expression is cumbersome and yields little insight, therefore we do not quote it here. 

Finally, we comment that at higher temperatures $\beta \Theta_0\lesssim 1$, the expressions given above for the susceptibility acquire additional terms reflecting the effect of Landau processes. These lead to further contributions to the imaginary parts of $\chi_\mathrm{pair}(q,\nu+\ii\epsilon)$, $\Omega(q,\nu+\ii\epsilon)$ and $B(q,\nu+\ii\epsilon)$, which play a similar role to the Beliaev contributions discussed above, except that they may be appreciable even at frequencies $|\nu|\ll \Theta_0$.